\documentclass[journal]{IEEEtran}
\usepackage{ifpdf}

\usepackage{cite}

\ifCLASSINFOpdf
    \usepackage[pdftex]{graphicx}
    \graphicspath{{./graphics}}
    \DeclareGraphicsExtensions{.pdf,.jpeg,.png}
\else
\fi
\usepackage{amsmath}
\usepackage{amsfonts}
\usepackage{algorithm}
\usepackage{algorithmic}

\usepackage{array}

\ifCLASSOPTIONcompsoc
    \usepackage[caption=false,font=normalsize,labelfont=sf,textfont=sf]{subfig}
\else
    \usepackage[caption=false,font=footnotesize]{subfig}
\fi
\usepackage{multirow}
\usepackage[table]{xcolor}
\usepackage{threeparttable}
\usepackage{makecell}

\usepackage{hyperref}
\usepackage{url}

\begin{document}
%
\title{From Large-scale Audio Tagging to Real-Time\\Explainable Emergency Vehicle Sirens Detection}
%
%
%

\author{Stefano~Giacomelli,~\IEEEmembership{Graduate Student Member,~IEEE,}
        Marco~Giordano,~\IEEEmembership{Student Member,~IEEE,}
        Claudia~Rinaldi,~\IEEEmembership{Member,~IEEE}
        and~Fabio~Graziosi,~\IEEEmembership{Member,~IEEE}
\thanks{S. Giacomelli, M. Giordano and F. Graziosi are with the Department of Information Engineering, Computer Science and Mathematics (DISIM), University of L'Aquila, L'Aquila, 67100, ITA e-mail: (stefano.giacomelli@graduate.univaq.it, marco.giordano3@graduate.univaq.it, fabio.graziosi@univaq.it).}
\thanks{C. Rinaldi is with the National Inter-University Consortium for Telecommunications (CNIT), University of L'Aquila, L'Aquila, 67100, ITA e-mail: (claudia.rinaldi@univaq.it).}
\thanks{Manuscript received MONTH XX, 20XX; revised MONTH XX, 20XX.}}

%
%

\markboth{IEEE/ACM TRANSACTIONS ON AUDIO, SPEECH, AND LANGUAGE PROCESSING,~Vol.~XX, No.~XX, Month~2025}%
{Shell \MakeLowercase{\textit{et al.}}: Bare Demo of IEEEtran.cls for IEEE Journals}
%



\maketitle

\begin{abstract}
Accurate recognition of Emergency Vehicle (EV) sirens is critical for the integration of intelligent transportation systems, smart city monitoring systems, and autonomous driving technologies. Modern automatic solutions are limited by the lack of large-scale, curated datasets and by the computational demands of state-of-the-art sound event detection models. This work introduces E2PANNs (Efficient/Emergency Pre-trained Audio Neural Networks), a lightweight Convolutional Neural Network architecture derived from the PANNs framework, specifically optimized for binary EV siren detection. Leveraging our dedicated subset of AudioSet (AudioSet-EV) we fine-tune and evaluate E2PANNs across multiple reference datasets and test its viability on embedded hardware. The experimental campaign includes ablation studies, cross-domain benchmarking, and real-time inference deployment on edge device. Interpretability analyses exploiting Guided Backpropagation and Score-CAM algorithms provide insights into the model’s internal representations and validate its ability to capture distinct spectro-temporal patterns associated with different types of EV sirens. Real-time performance is assessed through frame-wise and event-based detection metrics, as well as a detailed analysis of false positive activations. Results demonstrate that E2PANNs establish a new state-of-the-art in this research domain, with high computational efficiency, and suitability for edge-based audio monitoring and safety-critical applications.
\end{abstract}


%
\IEEEpeerreviewmaketitle

\section{Introduction}
\IEEEPARstart{E}{mergency} Vehicle (EV) siren recognition plays a crucial role in autonomous driving, smart city infrastructure, and urban acoustic monitoring. Accurate \textit{moving sirens} detection enhances traffic safety by enabling intelligent vehicle responses, real-time (RT) navigation adjustments, and automated driver alerts~\cite{5609617, 8058008, electronics11040510, 6936857}. Despite growing research interest and encouraging benchmark results, the development of robust Machine Learning (ML) solutions for this task remains challenging. Major obstacles include the lack of large-scale, curated datasets and the limited availability of lightweight yet reliable models suitable for deployment on resource-constrained devices. Although recent advances in Deep Learning (DL) have significantly improved performance in various audio recognition tasks — from General-Purpose Audio Tagging (GP-AT) to Sound Source Localization (SSL)~\cite{grumiaux_survey_2022} and Anomalous Sound Event Detection (SED)~\cite{c-sed_2} — few works have demonstrated effective real-world integration of such models for EV siren detection.

To address SED challenges for EVs~\cite{choudhury_review_2023}, a promising strategy involves leveraging pre-trained and high-performing classifiers from the GP-AT domain. By \textit{fine-tuning} these multi-objective models, specialized EV detectors can be derived and deployed to perform inference on short RT-acquired audio segments. This approach offers a practical trade-off between model complexity and efficiency, reducing reliance on recurrent NN architectures which, albeit effective in state-of-the-art (SoA) sound detectors, typically require substantial computational resources.

While Transformer-based architectures continue to gain traction~\cite{zaman_transformers_2025, tay_efficient_2023, chitty-venkata_survey_2023}, Convolutional Neural Networks (CNNs) remain a compelling choice for audio tasks due to their well-established reliability, reduced training complexity, and ease of hardware optimization~\cite{habib_optimization_2022}. Their architectural simplicity also facilitates deployment on low-power edge platforms.

Motivated by these considerations, this paper introduces E2PANNs (Efficient/Emergency Pre-trained Audio Neural Networks), a lightweight yet high-performance CNN-based architecture derived from the established PANNs framework~\cite{kong_panns_2020}. Our approach leverages fine-tuning strategies applied to a GP-AT model, optimizing it specifically for the binary detection of EV siren events. The fine-tuning process utilizes our publicly released \textit{AudioSet-EV} dataset~\cite{giacomelli_2025_14882314}, a curated domain-specific subset derived from AudioSet~\cite{gemmeke_audio_2017}, which ensures semantic consistency, balanced class representation, and comprehensive metadata annotations for reproducibility and benchmarking. Specifically, the contributions and objectives of this work, designed to bridge critical gaps identified in the current literature, are summarized as follows:
\begin{itemize}
    \item To design, fine-tune, and rigorously optimize E2PANNs as a specialized binary EV siren classifier, preserving its architectural structure but enhancing classification performances.
    \item To validate the public release of \textit{AudioSet-EV}~\cite{giacomelli_2025_14882314}, facilitating a standardized benchmark for comparative analysis in EV siren recognition tasks.
    \item To rigorously evaluate the practical applicability and robustness of the proposed pipeline through extensive ablation studies, comprehensive benchmarks on multiple public datasets, and embedded real-world deployment tests using a Raspberry Pi 5 setup, demonstrating clear improvements over SoA solutions in computational efficiency, representation quality, and detection accuracy.
    \item To demonstrate the feasibility of achieving RT EV siren detection without \textit{recurrent layers}, explicitly addressing and mitigating known computational overhead and latency challenges typically associated with these kind of models.
\end{itemize}

The remainder of this paper is organized as follows: Section~\ref{soa} provides a comprehensive review of the current SoA in acoustic EV siren recognition, covering algorithmic approaches, relevant datasets, and hardware-oriented embedding strategies. Section~\ref{preliminary_analysis} briefly reviews our preliminary analysis of the AudioSet taxonomy and dataset organization, motivating the construction of the ad-hoc EV-siren benchmark~\cite{giacomelli-preprint}, and highlighting key challenges in adapting PANNs models to resource-constrained frameworks. In Section~\ref{implementation}, we describe the implementation of our proposed model, detailing architectural modifications, fine-tuning procedures, optimization strategies, and deployment for RT inference on experimental embedded devices. Section~\ref{results} provides an in-depth evaluation of our model through ablation studies, hyperparameter search experiments, and benchmarking against standard baselines and datasets introduced in Section~\ref{soa}. Finally, Section~\ref{conclusions} summarizes our main contributions, discusses current limitations, and outlines directions for future research.

\section{State of the Art \& Related Works} \label{soa}
A growing body of research has addressed the problem of acoustic EV siren recognition, motivated by its relevance in safety-critical systems such as autonomous driving and smart mobility. In this section, we provide a structured review of the most relevant contributions in the field, with the aim of contextualizing our methodological choices and benchmarking strategies. We begin by analyzing recent developments in model architectures and detection strategies, focusing on their design principles, evaluation protocols, and reported performance. We then examine embedded-oriented implementations, with particular attention to solutions targeting RT inference under computational constraints. Finally, we outline the main datasets commonly adopted for training and evaluation. A more detailed data analysis of curated datasets and benchmarking tools is presented in a separate contribution~\cite{giacomelli-preprint}, which complements the present overview and directly informs the methodological framework introduced in the next sections.

\subsection{Emergency Vehicle Siren Recognition Models}
Acoustic EV siren detection has gained increasing attention due to its potential to enhance driver awareness and road safety. A recent survey~\cite{choudhury_review_2023} highlights the diversity of approaches in this domain, spanning from traditional Digital Signal Processing (DSP) techniques to advanced DL models, and underscores the challenges imposed by real-world noisy environments. In this section, we review notable contributions, focusing on how they address model design, data augmentation, and deployment constraints.

\begin{table*}[ht]
\centering
\caption{Summary of EV-siren recognition solutions}
\label{tab:ev_siren_comp}
\begin{tabular}{|p{2.6cm}|p{2.2cm}|p{2.5cm}|p{2.5cm}|p{2.5cm}|p{2.5cm}|}
\hline
\textbf{Authors} &
\textbf{Dataset \& Inputs} &
\textbf{Pre-processing} &
\textbf{Model / Algorithm} &
\textbf{Test Results} &
\textbf{Limitations}\\
\hline

Castorena et al.~\cite{castorena_safety-oriented_2024} &
Synthetic dataset (19,000 10s mixtures) &
Mel-spec, $\lambda$-mixup, patching &
CRNN (BiGRU), YOLOv5-based CNN &
Acc: 97\%, Inf.: 29\,ms (GPU), 226\,ms (CPU) &
No HW eval., synthetic only \\
\hline

Ramirez et al.~\cite{ramirez_siren_2022} &
London recordings (10,000 files) &
Spectrograms &
2D CNN &
Acc: 91\%, RT eval.: 48--71\% &
No inf. latency reported, overfitting risk \\
\hline

Shams et al.~\cite{shams_acoustic_2024} &
LSSiren (600 clips, 3s) &
MFCCs, pitch/time shift, noise inj. &
EfficientNet + 1D CNN + MHSA &
Acc: 100\%, Inf.: 209\,ms &
Small dataset, no HW eval. \\
\hline

Mittal et al.~\cite{mittal_acoustic_2023} &
AudioSet (10s) &
MFCCs &
Ensemble: FC, CNN, LSTM (voting) &
Acc: 98.7\%, Inf.: 1.5\,s (ensemble) &
High latency, no HW eval. \\
\hline

Carmel et al.~\cite{carmel_detection_2017} &
70 clips (30s), 100\,ms frames &
BPF, Hilbert, DWT, WPT, YIN, MFCCs &
ReliefF + SVM &
Acc: 98\% (10-fold) &
Tiny dataset, no complexity eval. \\
\hline

Tran et al.~\cite{tran_acoustic-based_2020} &
YouTube, Taiwan rec., ESC-50, US8K &
Raw + log-Mel (dual stream) &
SirenNet (WaveNet + CNN) &
Acc: 96.9–98.2\% (250\,ms input) &
No inf. latency, unclear dataset, no HW eval. \\
\hline

Shah et al.~\cite{shah_audio_2023} &
1675 clips from~\cite{tran_acoustic-based_2020} &
MFCCs &
CNN, LSTM, GRU (GRU best) &
Acc: 98.8\% &
No latency/HW analysis \\
\hline

Marchegiani et al.~\cite{marchegiani_listening_2022} &
Ad-hoc stereo dataset &
2ch gammatonegrams (2.5\,s) &
U-Net + FC + DoA regression &
Acc: 94\%, Err: 2.5° &
Offline only, stereo input required \\
\hline

Cantarini et al.~\cite{cantarini_few-shot_2022} &
Spoken Wiki, US8K, A3Siren &
Log-Mel + CNN embedding &
ProtoNet (few-shot learning) &
AuPRC: 86–91\% (w/ HPSS) &
No inf. latency/HW eval. \\
\hline

\end{tabular}
\end{table*}

Castorena et al.~\cite{castorena_safety-oriented_2024} proposed a hybrid detection pipeline combining DSP-based Mel-spectrogram extraction with two distinct DL architectures. The first is a Convolutional Recurrent Neural Network (CRNN) that integrates Bi-directional Gated Recurrent Units (Bi-GRU) and \textit{fully connected} (FC) layers to capture temporal dynamics. The second employs a \textit{You Only Look Once} (YOLOv5)-based convolutional framework with spatial pyramid pooling and multi-scale prediction heads, allowing direct localization of siren events within the spectrogram domain. Both models were trained on a purpose-built synthetic dataset simulating realistic driving conditions, composed of 19,000 ten-second audio mixtures of in-vehicle and free-field recordings. Data augmentation strategies included background noise injection, \textit{mixup}~\cite{zhang_mixup_2018}, and patch-based synthesis. Evaluation results showed that the YOLO-based model outperformed the CRNN baseline, achieving an average inference time of 29~ms on an NVIDIA RTX 3060Ti and 226~ms on a Cortex-A72 CPU. However, real-time applicability and embedded deployment considerations were not explicitly addressed.

Ramírez et al.~\cite{ramirez_siren_2022} proposed a CNN-based acoustic alert identification system designed to assist hearing-impaired individuals by detecting the presence and type of alert signals in noisy urban environments. Input audio is transformed into spectrogram representations, which are processed by Convolutional blocks [CNN layers $\rightarrow$ Rectified Linear Unit (ReLU) $\rightarrow$ pooling operations] to reduce feature dimensionality. The resulting feature maps are then passed to a FC classifier that outputs predictions over five categories: four distinct siren types and a control class representing background noise. A balanced dataset of 10,000 labeled spectrograms was constructed by combining three siren recordings per class with five different ambient noise samples, collected from London street recordings. The model achieved 97\% validation accuracy and 91\% test accuracy on this dataset, but the performance drop between validation and testing suggests potential overfitting to specific acoustic characteristics of the London urban environment. Continuous audio stream evaluations further revealed substantial variability in recognition rates (ranging from 48.5\% to 70.7\%), largely attributable to changes in real-world signal-to-noise ratios (SNRs).

Shams et al.~\cite{shams_acoustic_2024} proposed a hybrid framework for EV siren detection that integrates an EfficientNet backbone~\cite{pmlr-v97-tan19a} with one-dimensional CNN layers and Self-Attention mechanisms (SA-CNN). Input audio is segmented into 3-second clips and transformed into Mel-Frequency Cepstral Coefficients (MFCCs), which are used as input features. EfficientNet leverages depth-wise separable convolutions for compound scaling, balancing network depth, width, and resolution. The 1D CNN module is designed to extract temporal features, while SA layers~\cite{NIPS2017_3f5ee243} help capture long-range dependencies and improve discrimination between siren signals and background traffic noise. To train the model, the authors constructed a binary dataset comprising 600 manually assembled samples labeled as either \texttt{emergency vehicle} (including \texttt{ambulance} and \texttt{firetruck} sirens) or \texttt{traffic noise}. Data augmentation~\cite{electronics11223795} during training included pitch shifting, time stretching, and noise injection. The proposed system achieved high classification accuracy and demonstrated an average inference time of 209~ms per audio segment. However, due to the limited dataset size and absence of embedded testing, open questions remain regarding the model’s generalization to large-scale or real-world deployment scenarios.

Mittal et al.~\cite{mittal_acoustic_2023} proposed an ensemble-based framework for acoustic siren recognition using features derived from AudioSet~\cite{gemmeke_audio_2017}. Input clips are pre-processed to extract MFCCs, which are then fed into three independent NN architectures: an 8-layer FC network (FC-Net), a two-dimensional CNN with three convolutional layers and max-pooling, and a recurrent neural network (RNN) composed of five Long Short-Term Memory (LSTM) units. The final classification is obtained via static majority voting across the three model outputs. The ensemble achieved a test accuracy of 98.7\%, outperforming each individual model (FC-Net: 96.4\%, CNN: 92.4\%, RNN: 94.5\%). However, this performance gain comes at the cost of increased computational latency, with an average inference time of approximately 1.5 seconds per audio sample, compared to 0.061–0.151 seconds for the individual architectures.

Carmel et al.~\cite{carmel_detection_2017} presented a statistical ML approach for detecting alarm-related acoustic events — including sirens, alarm clocks, and fire alarms — under challenging environmental noise conditions. The pre-processing stage includes band-pass filtering (500–1500~Hz) and temporal envelope extraction via Hilbert transform. Feature extraction is performed every 100~ms with 50\% overlap, resulting in an effective detection delay of 200~ms. A comprehensive set of features is computed across three domains: time-domain features (e.g., pitch via the YIN algorithm~\cite{10.1121/1.1458024}, short-time energy, zero-crossing rate), frequency-domain features (MFCCs, spectral flux, roll-off, centroid, and flatness), and wavelet-domain features derived from both Discrete Wavelet Transform (DWT) and Wavelet Packet Transform (WPT). Feature selection is performed using the ReliefF algorithm~\cite{sikonja_theoretical_2003} with 35 nearest neighbors, retaining the most discriminative descriptors and discarding approximately 12\% of the original feature set. The authors compiled a custom dataset of 70 audio clips, each 30 seconds long, comprising 20 isolated alarm events, 15 alarm events embedded in noise, and 35 background noise samples. A Support Vector Machine (SVM) classifier trained on the selected features achieved 98\% accuracy using 10-fold cross-validation. Nonetheless, the study lacks a detailed analysis of computational complexity and does not explore the feasibility of real-time or embedded deployment, limiting its applicability in resource-constrained scenarios.

Tran et al.~\cite{tran_acoustic-based_2020} introduced \textit{SirenNet}, a dual-stream ensemble architecture designed for acoustic siren recognition in both in-vehicle and environmental settings. The system combines two parallel CNN-based branches: one stream (WaveNet) processes raw audio waveforms in an end-to-end manner, while the other (MLNet) operates on handcrafted representations, including MFCCs and log-mel spectrograms. The outputs of both branches are fused via softmax averaging, enabling the model to leverage complementary feature representations. The network is trained and evaluated on recordings of \texttt{ambulance}, \texttt{fire engine}, and \texttt{police car} sirens captured under various real-world conditions. However, the exact dataset used is not disclosed, limiting the reproducibility of the reported results. SirenNet achieves an accuracy of 98.24\% on full-length samples and maintains 96.89\% accuracy even when inputs are reduced to 250~ms. Nonetheless, the authors do not explicitly address continuous inference scenarios or assess the system’s performance in streaming or real-time applications.

Shah et al.~\cite{shah_audio_2023} explored a DL framework for the offline classification of emergency vehicle sirens using MFCCs as input features. The study compares three NN architectures — a CNN, a LSTM, and a GRU — each trained and evaluated independently. Among the three, the GRU-based model achieved the highest accuracy of 98.80\% in distinguishing between \texttt{ambulance}, \texttt{fire truck}, and \texttt{police car} sirens. The training/evaluation dataset comprises 1,675 audio clips of variable duration, containing real-world traffic noise and siren events, and is derived from the corpus presented in~\cite{tran_acoustic-based_2020}. While the experimental results are promising, the study does not discuss generalization to continuous or streaming inference, nor does it address deployment feasibility in real-time or embedded scenarios.

Marchegiani et al.~\cite{marchegiani_listening_2022} proposed a DL framework that adapts image segmentation techniques to time–frequency audio representations for siren detection and localization. A U-Net architecture is employed in a multi-task learning framework, where it simultaneously performs denoising of the input and classification of siren events via FC layers with Exponential Linear Unit (ELU) activation functions. In parallel, a separate convolutional regression network estimates the Direction of Arrival (DoA) by analyzing cross-correlations derived from the denoised two-channel gammatonegrams. Experimental results report an average classification accuracy of 94\% and a median localization error of 2.5°, using 2.5-second stereo audio segments under low signal-to-noise ratio (SNR) conditions (down to -40~dB). While the approach demonstrates strong performance, its architectural complexity and reliance on stereo input may pose challenges for real-time or embedded deployment.

Cantarini et al.~\cite{cantarini_few-shot_2022} proposed a workflow for ambulance siren detection based on \textit{prototypical networks}, a few-shot learning paradigm~\cite{snell2017prototypicalnetworksfewshotlearning}. Training is performed in an episodic setting, where each episode consists of a small \textit{support} set (C-way, K-shot) and a corresponding \textit{query} set. A CNN-based embedding function maps input log-mel spectrograms into a latent space, and class \textit{prototypes} are computed as the mean embeddings of support examples. During inference, query samples are classified by comparing their embeddings to the prototypes using a similarity-based distance metric. The authors used a combination of datasets including the Spoken Wikipedia Corpora~\cite{10.1007/s10579-017-9410-y}, UrbanSound8K~\cite{salamon_dataset_2014}, and A3Siren~\cite{noauthor_few-shot-emergency-siren-detection_nodate}, and additionally validated their approach using multi-microphone in-vehicle recordings. The model achieved an Area Under the Precision–Recall Curve (AuPRC) of 86\% on unfiltered data, which improved to 91\% when harmonic–percussive source separation filtering was applied. While the approach is well-suited for data-scarce scenarios, no explicit discussion is provided regarding its suitability for real-time or embedded deployments.

Table~\ref{tab:ev_siren_comp} summarizes the key findings of this literature review, which reveals a diverse and evolving landscape of methodologies for EV siren recognition. Earlier studies primarily relied on handcrafted features and conventional classifiers, while more recent contributions adopt a range of DL paradigms — including CNNs, RNNs, attention-based models, and few-shot learning architectures — exhibiting heterogeneous levels of complexity and performance. Despite these advances, several limitations remain pervasive: many approaches are constrained by small-scale or synthetic datasets, and only a minority explicitly address issues related to inference latency, continuous or streaming operation, and deployment on resource-constrained platforms. These recurring gaps highlight the pressing need for scalable, efficient, and reproducible solutions capable of operating reliably in real-world scenarios.

Only a limited number of studies have explored hardware-centric solutions for EV siren detection, often prioritizing simplicity and energy efficiency over scalability and generalization. Miyazaki et al.~\cite{miyazaki_ambulance_2013} implemented a detection method on a dsPIC microcontroller by applying the FFT twice, aiming to compensate for Doppler shifts. While the approach achieved robustness in frequency-domain characterization, it incurred excessive processing delays (up to 8 seconds per sample), limiting its applicability in real-time systems. Meucci et al.~\cite{meucci_real-time_2008} proposed another microcontroller-based system that exploits pitch tracking based on dual-tone detection, tailored to sirens centered at 392~Hz and 660~Hz. Although suitable for specific European siren types, the method lacks generalization to diverse acoustic profiles. Beritelli et al.~\cite{beritelli_automatic_2006} introduced a low-cost algorithm for hearing-impaired driver assistance using Linear Prediction Coefficients (LPCs), extracted via Levinson–Durbin recursion. A detection event is triggered when selected LPCs remain within defined tolerance thresholds over time. The system was optimized for deployment on TI TMS DSPs, but was prone to false positives under uncontrolled conditions. In a complementary line of work, Dobre et al.~\cite{dobre_low_2015,dobre_improved_2017,dobre_high-performance_2024} proposed analog low-power circuits for siren detection, validated through SPICE simulations. While these solutions showed promising accuracy in controlled scenarios, they were not tested on larger datasets nor implemented on real PCBs, raising concerns about their practical deployability.

\subsection{Emergency Vehicle Audio Datasets} \label{sec:dataset_soa}
The performance and generalizability of EV siren recognition systems are intrinsically tied to the quality, scale, and taxonomic clarity of the datasets used for training and evaluation. While recent model developments exhibit increasing architectural sophistication, their true effectiveness depends on the representativeness and annotation consistency of the underlying corpora. We now review the most widely adopted datasets in the field, highlighting their structural features, label granularity, and limitations. This analysis provides the necessary background to understand the evaluation criteria applied in subsequent sections and motivates the introduction of our curated dataset, \texttt{AudioSet-EV}~\cite{giacomelli_2025_14882314}. The following review briefly summarize a more extensive analysis about this topic, we provided in \cite{giacomelli-preprint}. 

\begin{table*}[ht]
\centering
\caption{Summary of EV-Benchmark Datasets}
\label{tab:comparison_table}
\tiny
\begin{tabular}{|c|>{\columncolor{gray!15}}c|c|c|c|c|c|}
\hline
\textbf{Test Dataset}        & \textbf{AudioSet-EV (\textit{ours})} & \textbf{ESC-50}                     & \textbf{SireNNet}                                                 & \textbf{LSSiren}        & \textbf{UrbanSound8K}                                          & \textbf{FSD50K} \\ \hline
\textbf{Samples}             & 1,404                       & $72\times5$                         & [4, 8, 16, 32, 66, 134, 268, 536, 957, 1,258]                     & 1,834                   & [873, 888, 925, 990, 936, 823, 838, 806, 816, 837]             & 3,948           \\ \hline
\textbf{\textit{Positives}}  & 752                         & $8\times5$                          & [2, 4, 8, 16, 33, 67, 134, 268, 536, 837]                         & 448.71                  & [86, 91, 119, 166, 71, 74, 77, 80, 82, 83]                     & 55              \\ \hline
\textbf{\textit{Negatives}}  & 652                         & $64\times5$                         & [2, 4, 8, 16, 33, 67, 134, 268, 421, 421]                         & 932                     & [787, 797, 806, 824, 865, 749, 761, 726, 734, 754]             & 3,893           \\ \hline
\textbf{Duration (min.)}      & 234                         & $6\times5$                          & [0.2, 0.4, 0.8, 1.6, 3.3, 6.7, 13.4, 26.8, 47.85, 62.9]           & 902                     & [58.2, 69.2, 61.67, 66, 62.4, 54.87, 55.87, 53.75, 54.4, 55.8] & 1,773.43        \\ \hline
\hline        
\textbf{Sample Dur. (s)}  & 10                          & 5                                   & 3                                                                 & variable                & variable                                                       & variable        \\ \hline
\textbf{Audio Chs}           & 1                           & 1                                   & 1                                                                 & 1                       & variable                                                       & 1               \\ \hline
\textbf{Sampling Rate}       & 32kHz                       & 16kHz                               & 44.1kHz                                                           & variable (48kHz*)       & variable                                                       & 44.1kHz         \\ \hline
\textbf{Test Strategy}       & 10\% split                  & $5\times$ cross-val folders         & $10\times$ varying-size splits                                    & \textit{unknown splits} & $10\times$ fold splits                                         & eval split      \\ \hline
\end{tabular}
\end{table*}

Only two corpora have been explicitly designed for this task. \textit{SireNNet}~\cite{shah_sirennet-emergency_2023} comprises approximately 400 three-second clips per class (\texttt{firetruck}, \texttt{ambulance}, \texttt{police}, \texttt{traffic}), recorded at 44.1kHz and annotated with both ambient and \textit{isolated} siren signals. Augmented variants isolate the siren component for cleaner training signals. While the dataset offers consistent label granularity and an evaluation-friendly format with increasing split sizes (from 4 to over 1,200 samples), its overall content duration remains limited and the diversity of recording sources is modest. Moreover, its largest test split, although balanced (837 positives, 421 negatives), includes multiple variants of the same clip (original vs.\ augmented), introducing potential redundancy.

\textit{LSSiren}~\cite{lssiren, asif_large-scale_2022}, by contrast, targets greater ecological validity. It provides 1,800 variable-length clips (3–15s) sampled at 48kHz, equally divided into \texttt{Ambulance} and \texttt{Road Noise} classes. Recordings were captured under diverse acoustic conditions (different microphone placements and source distances) to increase signal variability. However, its taxonomic resolution is natively binary, the original sampling protocol remains partially undocumented, and its test splitting strategy is undefined, complicating replicability.

In addition to these specialized corpora, several GP audio datasets include EV-siren–related samples, albeit in limited form. \textit{ESC-50}~\cite{piczak_esc_2015} consists of 2,000 mono clips (5s, 44.1kHz) across 50 mutually exclusive classes grouped into five broad categories. Only 40 clips are labeled as \texttt{siren}, and no further taxonomic disambiguation (vehicle type) is available. Its small test set (72 samples across five folds, $\sim$6 minutes per fold) and high imbalance ($\sim$1:8) may limit metric interpretability.

\textit{UrbanSound8K}~\cite{salamon_dataset_2014} expands coverage to 8,732 field-recorded clips (up to 4,s), classified into 10 urban categories including \texttt{siren}. It supports evaluation via a 10-fold structure, but the absence of hierarchical labels and limited metadata hinders in-depth taxonomic analysis.

\textit{FSD50K}~\cite{fonseca_fsd50k_2022}, the largest corpus in this group, contains over 51,000 mono clips with multi-label annotations aligned to the AudioSet ontology. Labels such as \texttt{siren}, \texttt{ambulance}, and \texttt{police car} are present, but occur sparsely and inconsistently. The evaluation set shows a severe imbalance (55 positives vs.\ 3,893 \textit{urban sound} negatives, $\sim$1:71).

To address these challenges, we introduced \textit{AudioSet-EV}~\cite{giacomelli_2025_14882314}, a semantically consistent and reproducible subset of AudioSet, built using our dedicated \textit{AudioSet-Tools} framework~\cite{giacomelli-preprint}. This dataset includes over 8,000 10-second mono clips at 32kHz, selected through label-aware filtering mechanisms (\texttt{Emergency vehicle}, \texttt{Siren}, \texttt{Ambulance (siren)}, \texttt{Police car (siren)}, \texttt{Fire engine, fire truck (siren)}), with hierarchical generalization and balancing routines. Its public test subset (10\%) contains 1,404 clips, evenly balanced across 752 positives and 652 negatives (234 minutes total duration). Unlike existing corpora, AudioSet-EV ensures: \emph{(1)} label consistency; \emph{(1)} transparent and traceable metadata; and \emph{(1)} evaluation-ready structure with customizable splits. Furthermore, leveraging the Standardized Audio Label Taxonomy (SALT)~\cite{stamatiadis_salt_2024} and its Python API~\cite{noauthor_saltpy-salt_nodate}, we harmonized its label mappings with all SoA EV benchmarks, enabling taxonomically aligned comparisons. 

All datasets were integrated into our model evaluation pipeline via tailored PyTorch/Lightning AI loaders~\cite{paszke_pytorch_2019, falcon_pytorchlightningpytorch-lightning_2020} and pre-processing routines, ensuring fair benchmarking across both structural and taxonomic dimensions~\cite{giacomelli-preprint}. Table~\ref{tab:comparison_table} summarizes key dataset properties and test-set statistics, including cardinality, content duration, class balance, and sampling strategy.

\section{Preliminary Analysis} \label{preliminary_analysis}
This section lays the groundwork for the specialized design and optimization of our EV siren recognition model, based on the gaps and challenges identified in Section \ref{soa}. We begin by examining two core strategies for the efficient deployment of CNNs, with a focus on \textit{pruning} techniques that reduce inference complexity for embedded and low-power platforms. We then introduce the Efficient Pre-trained Audio Neural Networks (EPANNs)~\cite{singh_e-panns_2023} framework, adopted as our baseline due to its favorable trade-off between performance and computational footprint. Finally, we present an initial statistical evaluation conducted on a preliminary version of AudioSet-EV, aimed at assessing the baseline model's readiness for the targeted EV siren classification task and guiding the subsequent optimization steps.

\subsection{CNNs Optimization for Embedded Deployment}
CNNs have achieved SoA performance in almost all audio-related tasks. However, the computational demands of top-performing models often hinder their deployment on resource-constrained embedded systems for RT applications. Optimizing NNs for such environments has therefore become an active research area, encompassing techniques like \textit{knowledge distillation}, architectural pruning \cite{ghimire_survey_2022, habib_optimization_2022}, and even dedicated hardware implementations (e.g.: FPGAs, ASICs) \cite{kang_survey_2024}. Given our experimental deployment targets, we will focus on two pruning methodologies that align closely with the EPANNs framework~\cite{singh_e-panns_2023}.

\subsubsection{Redundancy-Based Pruning} \label{svd_pruning}
Singh et al.~\cite{singh_svd-based_2020} introduced a pruning strategy based on Singular Value Decomposition (SVD), initially applied to 1D CNNs. The core idea is to evaluate the informational richness of each convolutional filter by analyzing the rank of its output representations across dataset classes. Filters producing \textit{low-rank} embeddings are deemed redundant, and can therefore be pruned to reduce inference cost with minimal impact on performance.

This is achieved by constructing a \textit{response matrix} for each convolutional filter, where rows represent dataset examples and columns represent the flattened activation outputs. Let there be $p$ input examples and $N$ filters, each producing embeddings of size $1 \times s$. After Z-scoring, a set $\mathcal{P}$ of $N$ response matrices is obtained, where each $x_z \in \mathcal{R}^{p \times s}$ corresponds to the outputs of the $z$-th filter across the dataset. SVD is then applied to each response matrix:
\[
x_z = U \Sigma V^T
\]
with $U \in \mathcal{R}^{p \times p}$ and $V \in \mathcal{R}^{s \times s}$ unitary matrices, and $\Sigma \in \mathcal{R}^{p \times s}$ containing the singular values on the diagonal.

To assess redundancy, the number of \textit{significant} singular values in $\Sigma$ is computed using a reference threshold $\Gamma$, such that $C \leq \Gamma \leq \min(p,s)$, where $C$ is the number of classes. Filters for which the number of significant singular values falls below $\Gamma$ are considered to contribute limited variation to the model’s representational capacity and are thus pruned. This method offers an \textit{adaptive data-driven} pruning strategy that avoids reliance on fixed thresholds or weight magnitudes. By tailoring the pruning process to the intrinsic statistical structure of learned feature maps, it enables significant computational savings while maintaining model accuracy.

\subsubsection{Passive Pruning via Operator Norm}
To further reduce model complexity without requiring access to training data, Singh et al.~\cite{singh_efficient_2023-1} proposed a \textit{data-independent} pruning technique based on the operator norm of convolutional filters. Unlike conventional strategies that rely on L1 or L2 weight magnitudes, this method evaluates how much each filter contributes to the CNN’s input–output transformation, thus enabling more principled pruning decisions.

Given the filter bank $K_l = [F^{l, 1}, F^{l, 2}, \dots, F^{l, n_l}]$ at the $l$-th convolutional layer, the method operates channel-wise: for each input channel $c$, a matrix $V^c \in \mathcal{R}^{n_l \times k_l^v}$ is constructed, where each row is a vectorized version of the $c$-th slice of a 2D filter and $k_l^v = k_l \times k_l$. SVD is then applied:
\[
V^c = U \Sigma V^T
\]
where $\Sigma$ contains singular values $\sigma_c$ representing how strongly each input direction is stretched. These directions are then normalized and used to compute a \textit{salience score} for each filter, capturing its relative importance in the transformation process. As a result, filters with lower scores are considered less informative and are removed — along with their associated feature maps — without observing the data. Despite being entirely data-free, this operator-norm approach has been shown to match or even outperform dataset-dependent methods in maintaining model accuracy while significantly reducing pruning time.

\subsection{Efficient Pre-trained Audio Neural Networks (E-PANNs)}
Building upon the redundancy-aware pruning strategies discussed above, Singh et al.~\cite{singh_e-panns_2023} introduced Efficient Pre-trained Audio Neural Networks (EPANNs), a compact yet highly effective variant of the PANNs CNN-14 architecture~\cite{kong_panns_2020}, originally trained on AudioSet~\cite{gemmeke_audio_2017}. Despite CNN-14's impressive performance in large-scale GP-AT tasks involving 527 multi-labeled classes, its large footprint — with over 81 million parameters and more than 50 billion multiply-accumulate operations (MACs) per inference on 10-second audio clips — severely limits its suitability for deployment on resource-constrained embedded devices.

To address these computational constraints, EPANNs leverages structured pruning specifically targeting the six deepest convolutional layers of CNN-14, collectively responsible for approximately 99\% of the total model parameters. By ranking and selectively removing filters using the operator norm-based pruning method (Section~\ref{svd_pruning}), Singh et al.~demonstrated that pruning 25\% and 50\% of these filters resulted in parameter reductions of 41\% and 70\%, respectively, while preserving or even enhancing mean Average Precision (mAP). Remarkably, EPANNs achieved a substantial 36\% reduction in MACs, outperforming the original CNN-14 baseline despite its significantly lower parameter count. Even at aggressive pruning levels (e.g., 75\%), the decrease in performance remained marginal, as detailed in Tables~\ref{tab:epanns_pruning} and~\ref{tab:panns_epanns}.

Motivated by these results, we adopt EPANNs as our baseline model over alternative compact CNN architectures such as MobileNet and EfficientNet, due to its proven capability to retain AudioSet classification accuracy even after extensive pruning. Additionally, EPANNs inherits the robust audio feature extraction abilities intrinsic to the PANNs architecture, making it particularly suitable as a starting point for specialized task-specific fine-tuning.

\begin{table}[t]
    \centering
    \caption{E-PANNs pruning results compared to CNN-14\\(AudioSet baseline)}
    \label{tab:epanns_pruning}
    \begin{tabular}{|c|c|c|c|}
    \hline
        \textbf{Filters} & \textbf{Parameter} & \textbf{MACs}  & \textbf{mAP} \\
        \textbf{Pruned}  & \textbf{Reduction} & \textbf{Reduction} & \textbf{(vs. CNN-14)}  \\
    \hline
        25\% & 41\% & 24\% & 0.442 $>$ 0.431 \\
        \rowcolor{lightgray} 50\% & 70\% & 36\% & 0.434 $>$ 0.431 \\
        75\% & 78\% & 46\% & 0.420 $<$ 0.431 \\
    \hline
    \end{tabular}
\end{table}

\begin{table}[t]
    \centering
    \caption{Memory and computational footprint comparison\\(rounded values)}
    \label{tab:panns_epanns}
    \begin{tabular}{|c|c|c|c|c|}
    \hline
        \multirow{2}{*}{\textbf{Model}} & \textbf{Total}        & \textbf{Trainable}  & \textbf{Weights} & \multirow{2}{*}{\textbf{FLOPs}} \\
                                       & \textbf{Parameters}   & \textbf{Parameters} & \textbf{Memory}  &                                \\
    \hline
        PANNs   & $81.8 \times 10^6$ & $80.7 \times 10^6$ & 312.1MB & $42.3 \times 10^9$ \\
        \hline
        EPANNs & $24.3 \times 10^6$ & $23.2 \times 10^6$ & 92.5MB  & $27.0 \times 10^9$ \\
    \hline
    \end{tabular}
\end{table}

\subsection{Preliminary Benchmark on AudioSet}
Before releasing the \textit{AudioSet-EV} dataset~\cite{giacomelli_2025_14882314}, we conducted a preliminary benchmark aimed at assessing the pre-trained EPANNs model in the context of EV-siren recognition. Using the \textit{AudioSet-Tools} processing suite~\cite{giacomelli-preprint}, we filtered AudioSet segments to construct the following test groups: 
\begin{itemize}
    \item \textit{Positives}, containing all samples tagged as \texttt{‘Emergency vehicle’};
    \item \textit{Negatives--Traffic}, including \texttt{‘Traffic noise, roadway noise’} and \texttt{‘Outside, urban or manmade’};
    \item \textit{Negatives--Vehicles}, with labels such as \texttt{‘Car passing by’}, \texttt{‘Truck’}, \texttt{‘Motorcycle’}, etc.;
    \item \textit{Negatives--Alarms}, including various non-EV alarm types (\texttt{‘Siren’}, \texttt{‘Car alarm’}, \texttt{‘Fire alarm’}, etc.).
\end{itemize}
To address potential label overlaps — common in multi-tag datasets — we removed any samples shared between \textit{Positives} and any \textit{Negative} group (see Table~\ref{tab:audioset-preliminary}).

\begin{table}
    \centering
    \caption{Preliminary AudioSet filtered segments statistics}
    \label{tab:audioset-preliminary}
    \begin{tabular}{|c|c|c|}
        \hline
        \textbf{Preliminary Set}       & \textbf{Samples}  & \textbf{Unique Labels} \\ \hline
        \textit{Positives} -- EV        & 5,730             & 74   \\
        \hline
        \textit{Negatives} -- Traffic   & 35,769            & 386  \\
        \textit{Negatives} -- Vehicles  & 38,305            & 309  \\
        \textit{Negatives} -- Alarms    & 10,815            & 154  \\ \hline
    \end{tabular}
\end{table}

Despite the large size of the negative sets, their semantic and acoustic contents vary significantly. Co-occurrence analysis [\href{https://github.com/StefanoGiacomelli/e2panns/blob/main/0_AudioSet_EPANNs_EV_Preliminary_Analysis.ipynb}{GitHub}] revealed that while \textit{Positives} mostly include labels strictly related to emergency contexts, the \textit{Traffic} subset includes over 21,000 samples (60\%) with speech, rather than vehicle sounds. This exemplify the semantic ambiguity and hierarchical inconsistency of AudioSet's taxonomy, as further analyzed in \cite{giacomelli-preprint}.

\begin{figure}
    \centering
    \includegraphics[width=1.0\linewidth]{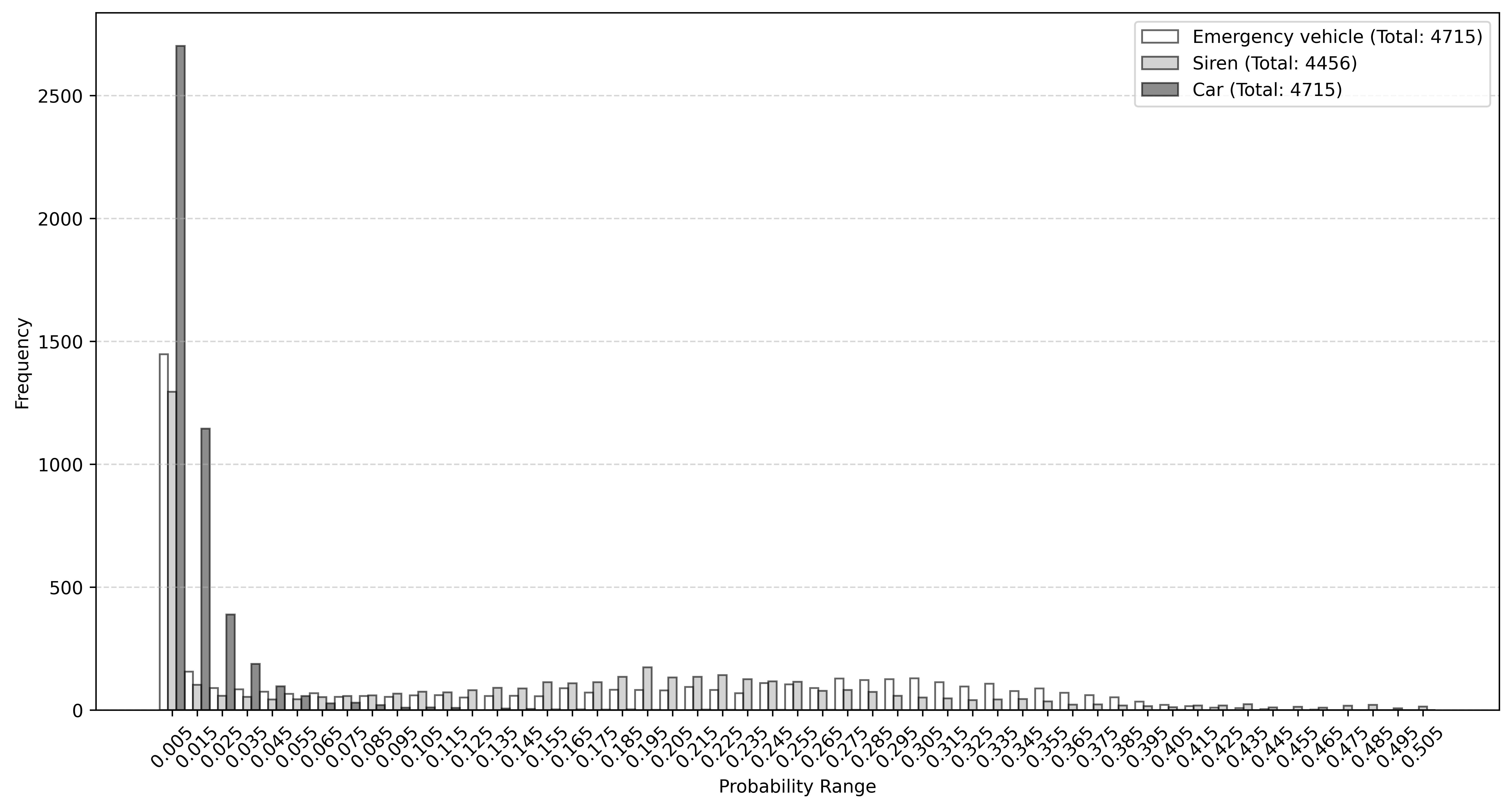}
    \caption{EPANNs output score (Class 322 - \texttt{‘Emergency vehicle’}) on \textit{Positives}: probability distribution over $[0, 0.5]$ interval (rest is zero).}
    \label{fig:positives_probs}
\end{figure}

We used these filtered subsets to evaluate the EPANNs model in its pre-trained configuration. Figure~\ref{fig:positives_probs} shows the inference scores on the \textit{Positives} group, revealing a skewed distribution below 0.5 — confirming the model's inability to recognize actual EV siren events (False Negatives). Performance on the \textit{Negatives--Alarms} group shows that although the model correctly classifies the majority of non-EV samples, it still produces 192 False Positives. Manual inspection of these samples indicates two likely causes: 
\emph{(1)} weak labeling (EV sirens may be present but not annotated), and \emph{(2)} confusion between acoustically similar alarms.

\begin{figure}
    \centering
    \includegraphics[width=1.0\linewidth]{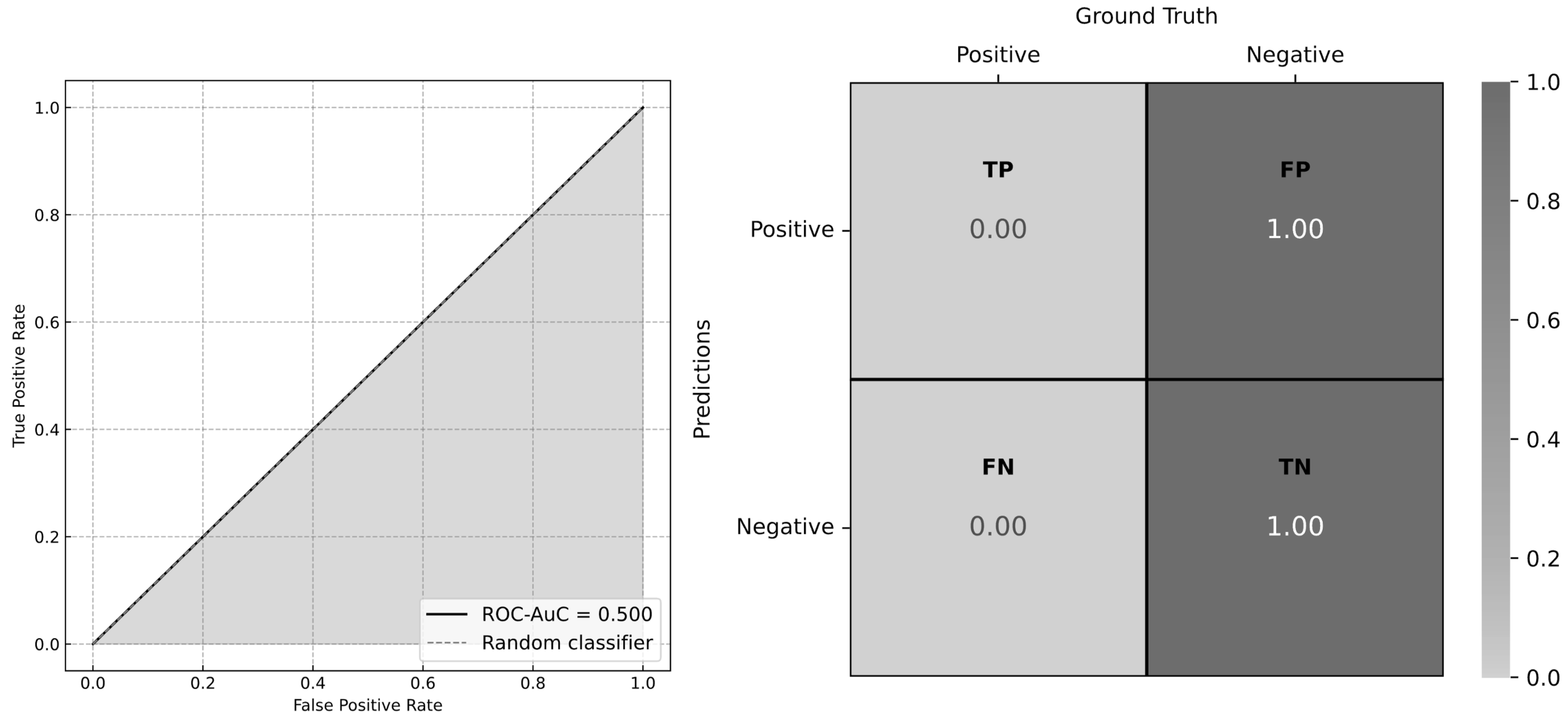}
    \caption{E-PANNs Confusion Matrix and ROC-AUC computed on preliminary AudioSet filtered samples.}
    \label{fig:epanns_pre_stats}
\end{figure}

The overall classification performance is summarized in Figure~\ref{fig:epanns_pre_stats}, showing a confusion matrix and ROC curve computed across all subsets. The results confirm that, despite inheriting pre-training on the full AudioSet, the EPANNs model behaves as a \textit{random classifier} with respect to EV-sirens. This outcome is expected, given that only $\sim$4.35\% of the full dataset intersects with our preliminary filtering (90,619 out of 2,084,320 samples), offering limited representation of emergency-related acoustic events.

In conclusion, our preliminary benchmark confirms that, despite the SoA performance achieved by EPANNs in large-scale GP-AT tasks, the model exhibits non-discriminative behavior when applied to the recognition of EV sirens. This limitation stems from the weak representation of EV-related classes in AudioSet, reinforcing the need for task-specific dataset curation and specialization~\cite{giacomelli-preprint}. Nonetheless, the EPANNs architectural efficiency and pruning-based modularity make it a compelling baseline for the downstream optimization strategies detailed in the next sections.

\section{Design \& Implementation} \label{implementation}
This section details the design and practical implementation of the proposed E2PANNs model pipeline. We first introduce the dataset management framework, covering data loading procedures, pre-processing steps, and partitioning strategies. We then describe the downstream fine-tuning of the selected EPANNs architecture, including training methodology and optimization processes. Next, we present the hyperparameter optimization strategy adopted to enhance model generalization. Finally, we describe the deployment of the optimized model on experimental edge hardware (Raspberry Pi 5), focusing on the development and validation of a RT inference system tailored to resource-constrained platforms. All experiments are discussed and critically analyzed in Section~\ref{results}.

\subsection{Datasets Management \& Standardization}
All datasets employed in our acoustic EV benchmarking pipeline~\cite{giacomelli-preprint} are managed through custom \texttt{PyTorch Dataset} classes and \texttt{Lightning DataModule} implementations. Each \texttt{Dataset} object loads audio files from disk, optionally applying re-sampling~\cite{brian_mcfee_bmcfeeresampy_2024}, and then enforces batch-wise input uniformity through fixed-length padding or truncation. A binary label is assigned to each sample (\texttt{siren}=1, \texttt{non\_siren}=0) using dataset-specific \textit{one-hot encoding} rules, as detailed in~\cite{giacomelli-preprint}. Corresponding \texttt{DataModule} instances handle \textit{train/validation/test} splits using either original metadata or configurable ratios, and provide the appropriate \texttt{DataLoader} interfaces for each phase.

\begin{table}[h]
    \centering
    \caption{Benchmarking Dataset partition sizes (in samples).}
    \label{tab:benchmark-partitions}
    \begin{tabular}{|c|c|c|c|}
    \hline
    \textbf{Dataset}                & \textbf{Train}    & \textbf{Validation}   & \textbf{Test} \\
    \hline
    AudioSet-EV                     & 11,220            & 1,402                 & 1,404         \\
    SireNNet                        & 1,340             & 167                   & 168           \\
    LSSiren                         & 1,467             & 183                   & 184           \\
    ESC50                           & 288               & 36                    & 36            \\
    UrbanSound8K                    & 6,985             & 873                   & 874           \\
    FSD50K                          & 14,421            & 3,606                 & 3,948         \\
    \hline
    \textbf{Unified-EV Dataset}     & 35,721            & 6,267                 & 6,614         \\
    \hline
    \end{tabular} 
\end{table}

\textit{AudioSet-EV} includes positive samples (\texttt{Emergency vehicle}) and negative urban traffic-related clips selected from our curated AudioSet redistribution~\cite{giacomelli_2025_14882314}. Each clip is mono, 32kHz, 10s long, and dataset splits follow an 80\%--10\%--10\% scheme. The same partitioning logic is applied to all other datasets not originally distributed with predefined split strategies (via custom implementations).

\textit{AudioSet-EV Augmented} extends our release version with time-domain data augmentation applied online during training, including random noise injection, polarity inversion, and temporal shifts~\cite{giacomelli-preprint}. At runtime, with probability 0.7, a block of augmentations is sequentially applied to each sample; each augmentation within the block is then probabilistically triggered based on an individual seeded mechanism. We deliberately chose not to apply additional spectral augmentations~\cite{electronics11223795}, as the EPANNs implementation already includes a built-in data augmentation layer that incorporates SpecAugment~\cite{park_specaugment_2019}, batch-wise $\lambda$-MixUp~\cite{zhang2018mixup}, and Patching~\cite{koutini_efficient_2022}. These operations are automatically triggered on input waveforms when the model is in \texttt{training} mode, according to the standard PyTorch workflow.

\textit{ESC-50}~\cite{piczak_esc_2015} is organized into 5 folds for cross-validation. For our purposes, we merge all clips into a single dataset, labeling ``siren'' samples as positives and all other urban or traffic-related classes as negatives. Files are re-sampled to 32kHz and zero-padded or truncated to 5s duration.

\textit{SireNNet}~\cite{shah_sirennet-emergency_2023} is scanned to extract positive (\texttt{ambulance}, \texttt{firetruck}, \texttt{police}) and negative (\texttt{traffic}) samples. Files are down-sampled from 44.1kHz, converted to mono, and normalized to 3s input tensors.

\textit{LSSiren}~\cite{asif_large-scale_2022} provides siren and road noise recordings, originally sampled at 48kHz. Files are converted to mono, down-sampled to 32kHz, and filtered by minimum duration (1s). Being a variable length dataset, a custom \texttt{collate} function pads inputs dynamically to the longest waveform in each batch.

\textit{UrbanSound8K}~\cite{salamon_dataset_2014} is merged across all 10 folds into a single dataset exploiting release metadata to assign binary labels. Due to variable file durations and sampling rates, a runtime \texttt{collate} function is used to perform re-sampling and zero-padding.

\textit{FSD50K}~\cite{fonseca_fsd50k_2022} is processed by scanning both \textit{development} and \textit{evaluation} folders. Positive and negative examples are extracted and labeled accordingly. The development split is shuffled and partitioned into \textit{train/validation} sets, while the \textit{evaluation} set is used as-is (post-label filtering) for testing. All files are converted to mono, re-sampled to 32kHz, and padded to a consistent tensor size.

For our fine-tuning experiments, we rely exclusively on \textit{AudioSet-EV} and its augmented variant. All other datasets serve only as benchmarking corpora for evaluating the generalization capability of the final model. Testing follows dataset-specific evaluation protocols, as summarized in Section~\ref{sec:dataset_soa}: \textit{AudioSet-EV} uses a 10\% hold-out test split, \textit{ESC-50} and \textit{UrbanSound8K} employ 5-fold and 10-fold cross-validation respectively, \textit{SireNNet} is evaluated over 10 randomized splits of increasing size, \textit{LSSiren} over the full dataset, and \textit{FSD50K} via its original evaluation partition.

Finally, we build a \textit{Unified-EV Dataset} by merging all benchmark sets into a single corpus, standardizing their input tensor dimensions to ensure compatibility. This merged dataset is used to evaluate the best-performing model under a unified \textit{transfer learning} scenario, enabling us to test cross-domain generalization across diverse urban and emergency acoustic contexts.

\subsection{Model Downstreaming \& Finetuning}
We adapt the EPANNs model --- originally trained for GP-AT over the 527 AudioSet classes~\cite{gemmeke_audio_2017} --- to the specific task of EV siren classifier, through a targeted model adaptation via fine-tuning. This involves converting the multi-label architecture into a binary classifier by isolating the prediction associated with the \texttt{Emergency vehicle} class (AudioSet index: 322). From the 527-dimensional pre-\textit{softmax} output vector, we extract the scalar activation corresponding to this class and apply a sigmoid function to produce a probabilistic output. A default threshold of 0.5 is used to classify the input as EV-siren (1) or non-EV-siren (0).

To enable specialization, we employ a selective finetuning strategy. The architecture is modified to optionally update only the final layers --- \texttt{fc1} and \texttt{fc\_audioset} in the original PyTorch implementation~\cite{giacomelli_stefanogiacomelliepanns_inference_2024} --- while keeping earlier feature extraction layers frozen. Alternatively, full fine-tuning of all layers can be enabled. The first configuration leverages transferable representations learned from the AudioSet corpus, limiting updates to higher-level features; the second allows a complete re-distribution of learning, enhancing specialization at the cost of increased training time.

The loss function is reformulated as Binary Cross-Entropy (BCE) in its log-likelihood form:

\begin{equation}
    \mathcal{L}(x_n,y_n) = -y_n\log(x_n) + (1 - y_n)\log(1 - x_n)
\end{equation}

where $x_n$ is the predicted probability for the $n$-th input sample, and $y_n \in \{0,1\}$ is the binary ground truth from the \textit{AudioSet-EV} dataset. The model prediction loss is computed using PyTorch \texttt{BCEWithLogitsLoss()}, integrating the sigmoid activation directly within the loss function for numerical stability and efficient optimization.

\begin{figure}
    \centering
    \includegraphics[width=1\linewidth]{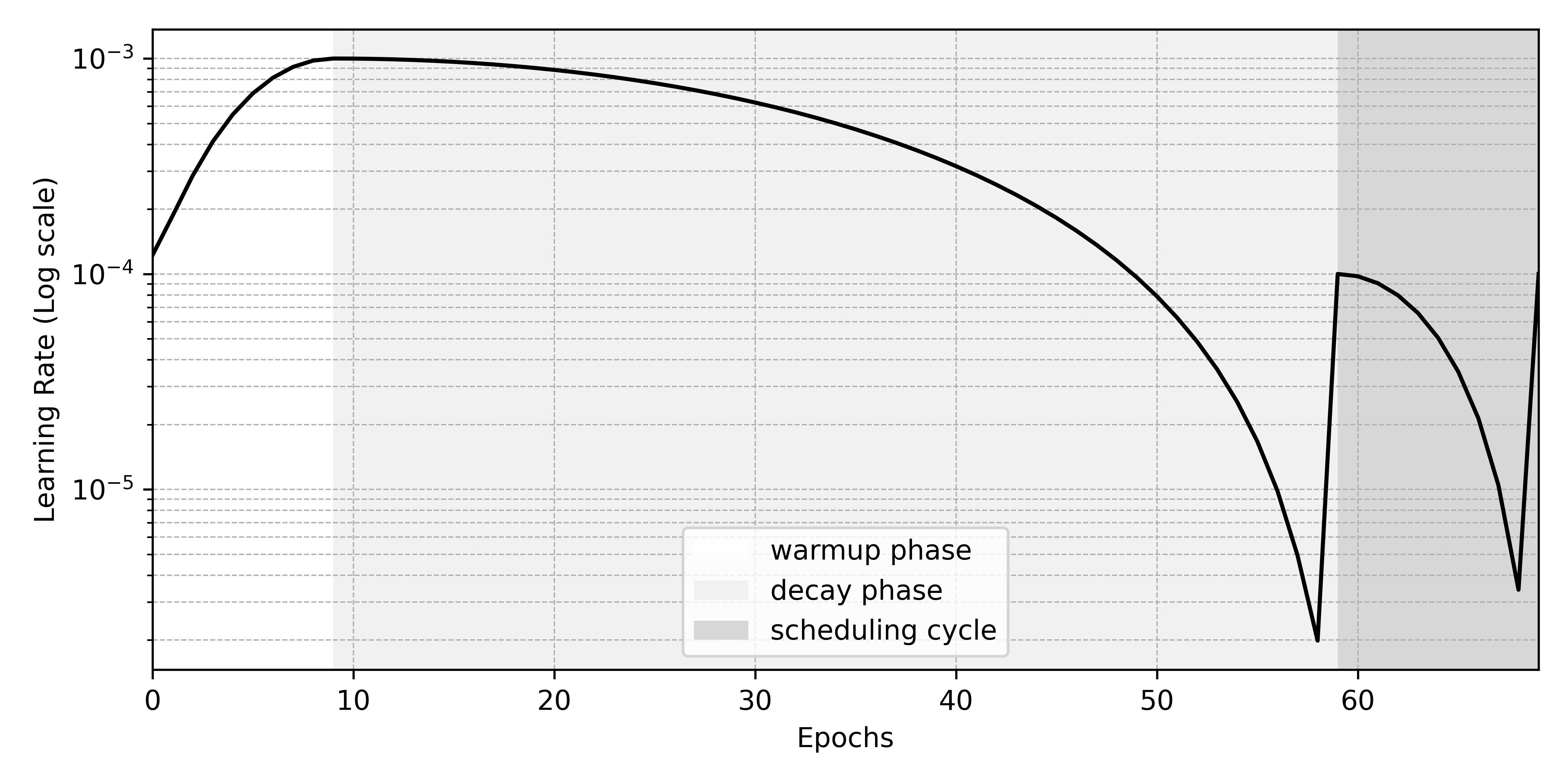}
    \caption{\textit{Cyclic Cosine Annealing Learning Rate scheduler}: a detail of the initial \textit{warm-up}, \textit{main decay}, and first restart cycle.}
    \label{fig:scheduler_zoom}
\end{figure}

Optimization follows the PANNs training protocol~\cite{kong_panns_2020}, using the Adam optimizer~\cite{kingma2017adammethodstochasticoptimization} with weight decay $\lambda$, and a cyclic cosine annealing learning rate scheduler~\cite{athiwaratkun2019consistentexplanationsunlabeleddata, loshchilov2017sgdrstochasticgradientdescent}. The scheduler includes warm-up epochs ($T_{\text{warm-up}}$) and periodic restarts ($T_{\text{cycle}}$), modulating the learning rate $\eta_t$ between $\eta_{\min}$ and $\eta_{\max}$, with $\eta_{\text{init}}$ as the cycle restart value. Figure~\ref{fig:scheduler_zoom} illustrates the scheduler behavior and a detailed pseudocode of the training procedure is shown in Algorithm~\ref{optimizer_alg}.

\begin{algorithm}
\caption{Adam w. Cyclic Cosine Annealing LR}
\footnotesize
\begin{algorithmic}[1]
\label{optimizer_alg}
\STATE Init model parameters: $\theta_0$
\STATE Init moment vectors: $m_0 \gets 0$, $v_0 \gets 0$
\STATE Init step counter: $t \gets 0$
\STATE Set optimizer Hyperparameters: 
    \begin{itemize}
        \item $\beta_1$: exponential decay rate for $1^{st}$ moment estimates 
        \item $\beta_2$: exponential decay rate for $2^{nd}$ moment estimates
        \item $\epsilon$: $10^{-8}$ (offset)
        \item $\lambda$: weight decay (regularization) coefficient
    \end{itemize}
\STATE Set scheduler Hyper-Parameters:
    \begin{itemize}
        \item $T_{\text{cycle}}$: total steps in one cycle,
        \item $T_{\text{warm-up}}$: steps for warm-up,
        \item $\eta_{\text{init}}$: learning rate at cycle restart,
        \item $\eta_{\max}$: peak learning rate after warm-up,
        \item $\eta_{\min}$: minimum learning rate at cycle end.
    \end{itemize}
  
\FOR{epoch \textbf{in \textit{epochs}}}
  \FOR{mini-batch \textbf{in \textit{dataset}}}
    \STATE $t \gets t + 1$
    
    \STATE \textit{Compute gradient ------------------------------------------------------------------} 
    \[g_t = \nabla_\theta f(\theta_t) + \lambda\, \theta_t\]
    
    \STATE $m_t \gets \beta_1\, m_{t-1} + (1-\beta_1)\, g_t$
    \vspace{0.5em}
    \STATE $v_t \gets \beta_2\, v_{t-1} + (1-\beta_2)\, g_t^2$
    \vspace{0.8em}
    
    \STATE $\hat{m}_t \gets \dfrac{m_t}{1-\beta_1^t}$, \quad $\hat{v}_t \gets \dfrac{v_t}{1-\beta_2^t}$
    \vspace{1.2em}
    \STATE \textit{Compute LR for current step ---------------------------------------------------}
    \STATE  $t_{\text{cycle}} = t \bmod(T_{\text{cycle}})$
    \vspace{0.8em}
    \IF{$t_{\text{cycle}} \leq T_{\text{warm-up}}$}
      \STATE $\eta_t \gets \eta_{\text{init}} + \dfrac{t_{\text{cycle}}}{T_{\text{warm-up}}}\, (\eta_{\max} - \eta_{\text{init}})$
    \ELSE
        \STATE $\eta_t \gets \eta_{\min} + \textstyle\frac{1}{2} (\eta_{\max} - \eta_{\min}) \left[ 1 + \cos\!\left(\pi\, \textstyle\frac{t_{\text{cycle}} - T_{\text{warm-up}}}{T_{\text{cycle}} - T_{\text{warm-up}}}\right) \right]$
    \ENDIF
    \STATE $\theta_{t+1} \gets \theta_t - \eta_t\, \dfrac{\hat{m}_t}{\sqrt{\hat{v}_t} + \epsilon}$
  \ENDFOR
\ENDFOR
\end{algorithmic}
\end{algorithm}

To mitigate overfitting and promote generalization, we apply \textit{early stopping}~\cite{Prechelt2012}, monitoring validation accuracy after each epoch. Training is halted when no improvement is observed for a given \textit{patience} window. This is implemented as a PyTorch Lightning~\cite{falcon_pytorchlightningpytorch-lightning_2020} \texttt{callback} function, with \texttt{mode="max"} to preserve configurations with highest validation accuracy. Additionally, model checkpoints are saved at peak performance to ensure best-inference reproducibility.

Model performance is evaluated using a comprehensive suite of metrics: Accuracy, Precision, Recall, F1-score, Specificity, Area Under the ROC Curve (AuROC), Area Under the Precision-Recall Curve (AuPRC), Matthews Correlation Coefficient (MCC), and the weighted $F_\beta$-score ($\beta = 0.8$), computed using Torchmetrics~\cite{detlefsen_torchmetrics_2022} and custom Python routines. Training statistics are logged in pseudo real time via \texttt{TensorBoard}, leveraging native Lightning AI integrations.

\subsection{Hyperparameters Search \& Finetuning Experiments}
The optimization of NNs hyperparameters is a critical process that directly affects training convergence and generalization performance. To identify an effective configuration, we employed a structured yet computationally efficient approximation of the \textit{grid search} method~\cite{10733621}. This heuristic-informed approach integrates hyperparameters space sub-sampling at strategically selected points, leveraging prior knowledge from HP configurations for PANNs and E-PANNs. Instead of exhaustive enumeration, the search focuses on refining known \textit{sweet spots}, allowing for targeted exploration. Previous works~\cite{10704155} has shown that this method provides a favorable trade-off between search efficiency and model performance. The resulting search space is reported in Table~\ref{tab:hparams}.

\begin{table}[h]
    \centering
    \caption{E2PANNs Hyperparameters Search Space}
    \label{tab:hparams}
    \begin{tabular}{|c|c|}
        \hline
        \textbf{Hyperparameters} & \textbf{Possible Values} \\
        \hline
        Training Mode & \texttt{overall, last layers} \\
        Batch Size & 16, 32 \\
        $\eta_{\max}$ & $10^{-5}$, $10^{-4}$, $10^{-3}$, $10^{-2}$ \\
        Binary Threshold & 0.5, 0.6, 0.7 \\
        \hline
    \end{tabular}
\end{table}

Certain hyperparameters were fixed \textit{a priòri} based on prior optimization studies and did not require further tuning, namely: weight decay $\lambda = 10^{-6}$ and minimum learning rate $\eta_{\min} = 10^{-6}$.

A total of 48 training/validation experiments (one per combination) were executed over a maximum of 50 epochs each, using an NVIDIA RTX A6000 GPU. A validation accuracy-based early stopping mechanism with a patience of 15 epochs was employed. Total runtime was approximately one day. Performance logs for each run are publicly available on our \href{https://github.com/StefanoGiacomelli/e2panns/blob/main/0_finetuning_results/hp_search_results.csv}{GitHub}. This process enabled the identification of an optimal model configuration without the prohibitive cost of exhaustive search. The resulting configuration was used in subsequent finetuning experiments. To further refine results, a potential \textit{local search} could be performed by perturbing the top-performing configurations within constrained margins, enhancing convergence robustness.

We then proceeded with the actual finetuning steps. The optimal hyperparameters and model design strategies were progressively tested (\textit{ablation studies}) and evaluated through a standard \textit{train/dev/test} split (80\% / 10\% / 10\%). Table~\ref{tab:finetuning_stages} summarizes the different finetuning stages conducted. For each session, detailed metric logs and console outputs were recorded, and results will be benchmarked and discussed in Section~\ref{results}.

\newcommand{\NA}{\cellcolor{black!70}}
{
\renewcommand{\arraystretch}{1.3}
\begin{table*}[h]
    \centering
    \begin{threeparttable}
    \caption{E2PANNs ablation stuies on AudioSet-EV}
    \scriptsize
    \begin{tabular}{|c|c|c|c|c|c|c|}
        \hline
        \multirow{2}{*}{\textbf{Experiment Stage Focus}}  & last layers finetuning                    & overall training                                  & $\eta$ exp. decay rate            & cosine $\eta$                             & dataset                       & fixed $\eta$                  \\
                                                    & $\eta$ exp. decay rate                    & $\eta$ exp. decay rate                            & (warm-up)                          & scheduler                                 & augmentation                  & dataset augmentation          \\
        \hline
        \noalign{\vspace{1em}}  
        \hline
        \textbf{Hyper-Parameters} & \textbf{Stage 1} & \textbf{Stage 2} & \textbf{Stage 3} & \textbf{Stage 4} & \textbf{Stage 5} & \textbf{Stage 6}\\
        \hline
        Batch Size                                  & \multicolumn{6}{c|}{32} \\
        \hline
        Training Mode                               & \cellcolor{gray!30} \textit{last layers}  & \cellcolor{gray!30} \textit{overall architecture} & \textit{overall architecture}     & \textit{overall architecture}             & \textit{overall architecture} & \textit{overall architecture} \\
        \hline
        Binary Threshold                            & 0.5                                       & 0.5                                               & 0.5                               & 0.5                                       & 0.5                           & 0.5                           \\
        \hline
        $\eta_{\text{init}}$                        & $10^{-6}$                                 & $10^{-6}$                                         & $10^{-6}$                         & \cellcolor{gray!30} $10^{-5}$             & $10^{-5}$                     & \cellcolor{gray!30} $10^{-4}$ \\
        \hline
        $\eta_{\max}$                               & $10^{-3}$                                 & $10^{-3}$                                         & $10^{-3}$                         & \cellcolor{gray!30} $10^{-3}$ ($10^{-4}$) & $10^{-3}$ ($10^{-4}$)         & \NA                           \\
        \hline
        $\eta_{\min}$                               & \NA                                       & \NA                                               & \cellcolor{gray!30} $10^{-6}$     & $10^{-6}$                                 & $10^{-6}$                     & \NA                           \\
        \hline
        $\lambda$                                   & $10^{-6}$                                 & $10^{-6}$                                         & $10^{-6}$                         & $10^{-6}$                                 & $10^{-6}$                     & $10^{-6}$                     \\
        \hline
        $T_{\max}$                                  & 10                                        & 10                                                & \cellcolor{gray!30} 100           & 100                                       & 100                           & \NA                           \\
        \hline
        $T_{\text{warm-up}}$                         & \NA                                       & \NA                                               & \cellcolor{gray!30} 5 Epochs      & 5 Epochs                                  & 5 Epochs                      & \NA                           \\
        \hline
        aug\_prob                                   & \NA                                       & \NA                                               & \NA                               & \NA                                       & \cellcolor{gray!30} 0.7       & \cellcolor{gray!30} 0.5       \\
        \hline
        \noalign{\vspace{1em}}  
        \hline
        Epochs                                      & 100                                       & 100                                               & 1000                              & 1000                                      & 1000                          & 1000                          \\
        \hline
        Early Stopping Patience                     & 33                                        & 33                                                & 50                                & 150                                       & 150                           & 100                           \\
        \hline
        \textbf{Test Accuracy}                      & 84.19\%                                   & 95.87\%                                           & \textbf{97.15\%}                  & 96.72\%                                   & 91.52\%                       & 91.81\%                       \\ 
        \hline
    \end{tabular}
    \begin{tablenotes}
    \centering
      \footnotesize
      \item Legend: \textcolor{darkgray}{dark grey} -- hyperparameters not applied, \textcolor{lightgray}{light grey} -- hyperparameters introduced or changed, \textbf{bold} -- best values.
    \end{tablenotes}
    \end{threeparttable}
    \label{tab:finetuning_stages}
\end{table*}
}

\subsection{Hardware Deployment for Real-Time Siren Detection} \label{subsec:edge}
Before deploying our E2PANNs model on the experimental target platform (Raspberry Pi 5) for RT EV-SED experiments, we evaluated its ability to perform inference on small input sizes. The goal was to reduce computational complexity by leveraging the finetuned EPANNs to perform periodic inference on short, variable-length segments of streaming audio data. This approach aims to avoid recurrent layers — common in SoA SED systems — which introduce substantial computational and memory overhead.

To determine the minimum viable input size, we implemented a binary search algorithm~\cite{DBLP:journals/corr/ChadhaMM14} to identify the smallest input tensor that produces a valid model output. The search space ranged from 1 to 320,000 samples (i.e.: 10s at 32kHz). At each step, the algorithm evaluates the midpoint duration, generating a corresponding audio tensor. If the model processes the input without runtime exceptions and returns a numerically valid output, the upper bound is updated to $midpoint - 1$; otherwise, the lower bound is set to $midpoint + 1$. This procedure continues until the bounds converge. After 19 iterations (10s), the minimum valid input size was found to be 9919 samples (approximately 310ms at 32kHz).

\begin{figure}
    \centering
    \includegraphics[width=0.95\linewidth]{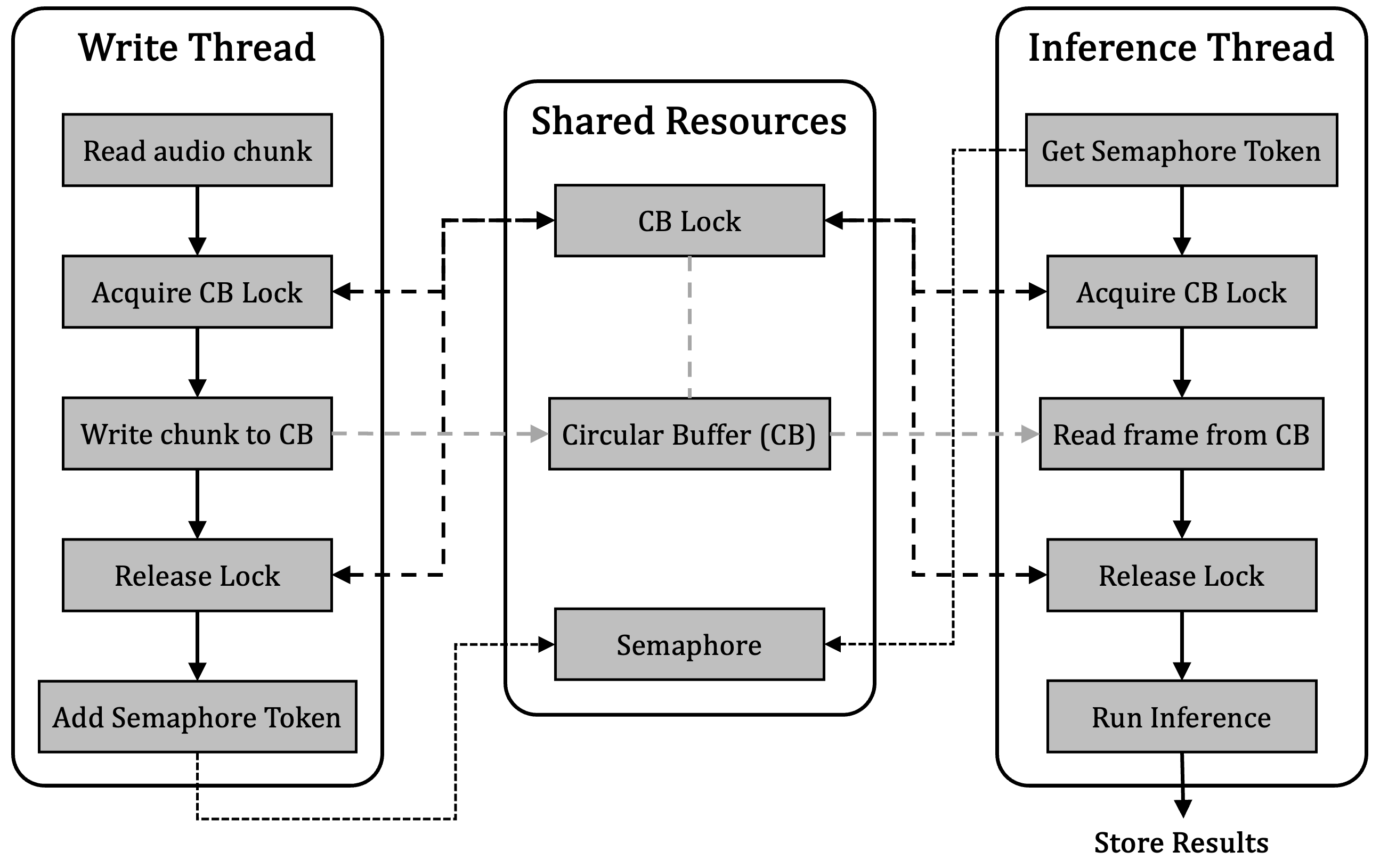}
    \caption{E2PANNs RT inference threading system running on Raspberry Pi 5.}
    \label{fig:RT_system}
\end{figure}

To support continuous EV siren detection on edge hardware, we designed a RT inference system tailored for the Raspberry Pi 5. The system (Figure~\ref{fig:RT_system}) adopts a multi-threaded architecture with explicit concurrency control, ensuring non-blocking and low-latency operation. Its core component is a \texttt{CircularBuffer} acting as shared memory between a \textit{producer} thread (writing audio data) and a \textit{consumer} thread (performing inference). Synchronization is enforced using Python’s \texttt{threading.Lock} and \texttt{threading.Semaphore} primitives. To ensure both simulation and live operational setup support, we integrated a RaspiAudio Ultra++ ADC/DAC audio board into our Raspberry Pi 5 platform. This board provides high-fidelity audio input with low-latency access to Advanced Linux Sound Architecture (ALSA) streams and onboard microphone array support.

The audio \texttt{Write Thread} (producer) emulates RT audio input by periodically writing short chunks from a source file or input buffer to the \texttt{CircularBuffer}. Thread-safe access is guaranteed by acquiring a \texttt{Lock} during buffer updates. Once data is written, a \texttt{Semaphore} token is released, signaling that new content is available for reading.

The \texttt{CircularBuffer} is a pre-allocated, fixed-size memory structure supporting wrapping reads and writes. It maintains a sliding window of the most recent samples, enabling continuous streaming behavior. Read/write operations are guarded by a \texttt{Lock}, while the \texttt{Semaphore} ensures reads occur only after new data has arrived.

The input \texttt{Frame Provider} module, embedded in the inference thread, manages frame extraction. Before reading, it acquires a \texttt{Semaphore} token (blocking if necessary), then locks the buffer to extract a valid audio frame. Frame length is dynamically adjusted based on the most recent model output: starting from the minimum valid size, the duration is increased whenever the output probability exceeds a predefined threshold (default 0.6). This adaptive mechanism progressively expands the temporal context during likely positive detections, enhancing robustness while limiting computational overhead when EV-like characteristics are absent. The frame duration is bounded by a configurable maximum.

The \texttt{Inference Thread} (consumer) runs concurrently, querying the \texttt{Frame Provider} for the next input and passing it to E2PANNs. The resulting probability is appended to a shared output list for decision logic. A core component of the post-processing stage is the \textit{event decision} state machine, which determines whether a positive siren detection event should be issued. This logic builds upon two key mechanisms: a fixed-size (3 consecutive frames) moving average filter applied to inference probabilities to suppress transient noise and reduce false positives; and the \textit{consecutive validation} mechanism which support positive events detection only if a minimum number of consecutive frames (each exceeding a configurable probability threshold) are observed. The inference loop terminates upon receiving a shutdown signal via a \texttt{threading.Event}.

The real-time engine is extended with a WebSocket-based interface for remote monitoring and control, exposing a live stream of classification probabilities, detection flags, and diagnostic metadata. Its frontend, accessible via browser or terminal client, allows operators to inspect system behavior during field deployment or controlled playback scenarios. The implementation achieves responsive and low-overhead inference while preserving data consistency. The adaptive frame-length mechanism complements the model's efficiency, allowing longer analysis windows only when acoustically justified. Source code and other runtime analysis details are available in a dedicated publication\cite{giordano_is2} with repository [\href{https://github.com/marco-giordano/e2panns-rbpi5}{GitHub}].

\section{Results Discussion} \label{results}
This section provides an in-depth analysis of the experimental outcomes derived from the methods and implementations discussed earlier. The results are structured along four main axes: hyperparameter optimization, multi-stage model fine-tuning, cross-dataset generalization via transfer learning, and deployment feasibility on embedded hardware.

We begin with the structured hyperparameter search, which guided the selection of training configurations that improved both convergence speed and model stability. We then examine the effects of a \textit{multi-stage fine-tuning} process on the AudioSet-EV dataset. This approach combines selective architectural adaptation with dynamic learning rate scheduling, yielding robust improvements in EV siren detection performance. These findings underscore the value of balancing parameter reuse and task-specific specialization, especially when dealing with noisy or sparsely labeled acoustic data~\cite{giacomelli-preprint}.

Next, we assess the generalization capabilities of the optimized E2PANNs model through a \textit{unified transfer learning} setup. By evaluating performance across our EV-benchmark (Section~\ref{sec:dataset_soa}), we demonstrate that the model consistently achieves SoA results. This confirms the versatility of our approach across diverse acoustic environments beyond AudioSet distributions.

Before turning to RT deployment, we present a brief architectural inspection of the classifier’s decision-making behavior to enhance its black box interpretability~\cite{molnar2025, 9369420}. We employ a suite of eXplainable AI (XAI) post-hoc techniques — Vanilla Backpropagation Saliency Maps, Grad-CAM, and Score-CAM — adapted for time–frequency inputs. These visualizations help identify which spectral regions most influence the model's activations, offering valuable insight into its internal reasoning processes and highlighting its focus on siren-relevant spectral patterns.

Finally, we report on embedded inference tests conducted on the Raspberry Pi 5. These confirm that our model supports low-latency EV detection while operating under tight computational constraints. The use of an adaptive frame-length strategy — modulating input duration based on classification confidence — allows the system to dynamically balance accuracy and resource consumption. Quantitative and qualitative results for each experimental stage follow in the remainder of this section.

\subsection{Hyperparameters Search \& Optimization}

\begin{figure}
    \centering
    \includegraphics[width=1.0\linewidth]{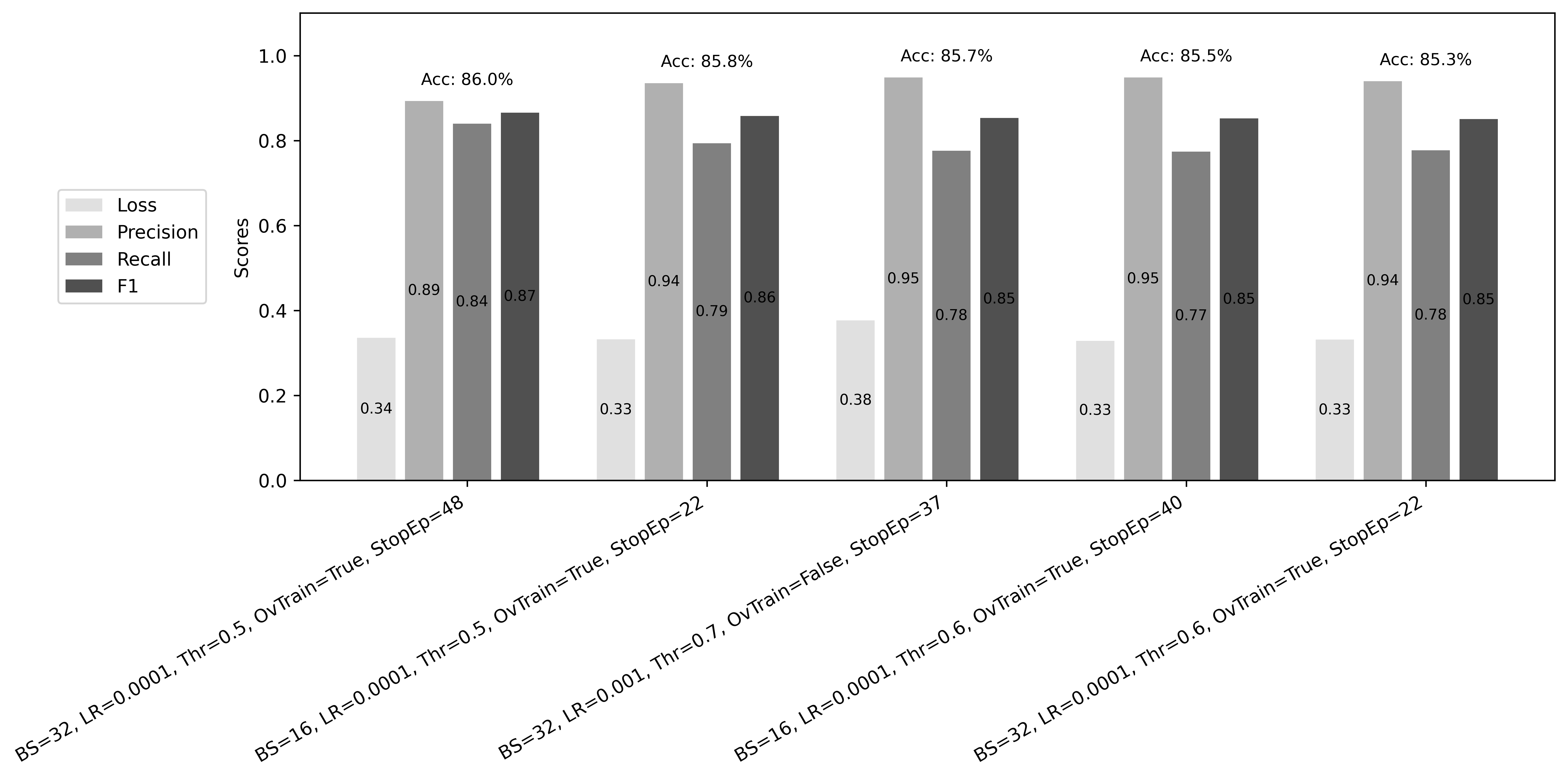}
    \caption{E-PANNs Hyperparameters Search results (Top-5 validation accuracy configurations): \texttt{BS}=Batch size, \texttt{LR}=Learning Rate, \texttt{Thr}=Binary Threshold, \texttt{OvTrain}=Overall Training mode, \texttt{StopEp}=Early Stopping Epoch}
    \label{fig:hp_search}
\end{figure}

We conducted a heuristic-informed hyperparameter search aimed at optimizing training stability and detection performance. Validation metrics for each configuration (Table~\ref{tab:hparams}) were collected and analyzed from the experiment logs available in our repository~[\href{https://github.com/StefanoGiacomelli/e2panns/tree/main/0_finetuning_results}{GitHub}]. Our analysis focuses on the effect of learning rate ($\eta$), batch size (BS), and binary classification threshold on validation accuracy, convergence speed, and detection trade-offs.

\begin{figure}
    \centering
    \includegraphics[width=1\linewidth]{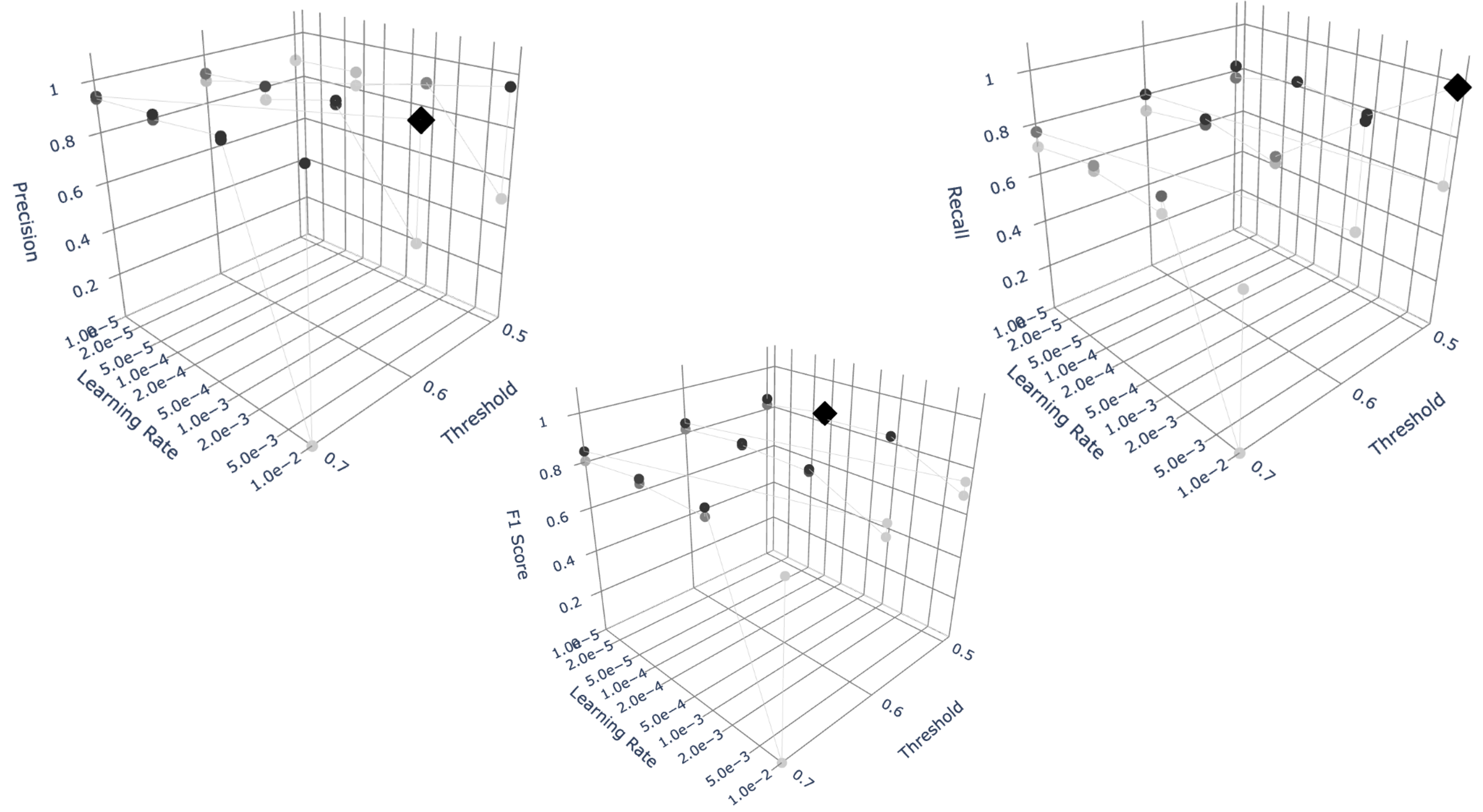}
    \caption{Precision, Recall, and $F_1$ score for Search configurations (BS = 32). \textcolor{darkgray}{Darker} point = higher scores, \textcolor{lightgray}{lighter} = lower scores, \textbf{bold diamond} = best score.}
    \label{fig:prec-recall-f1}
\end{figure}

Among the learning rates explored, intermediate values ($10^{-4}$ and $10^{-3}$) consistently led to the best outcomes (Figures~\ref{fig:hp_search},~\ref{fig:prec-recall-f1}), achieving validation accuracies up to 86\% and avoiding both underfitting and divergence. Extremely low rates ($10^{-5}$) caused slow convergence, while high rates ($10^{-2}$) introduced instability, resulting in validation losses exceeding 50. These observations confirm that excessively small $\eta$ may suppress gradient updates, while excessively large values may overshoot optimal minima.

Variations in the classification threshold strongly influenced the inverse correlation between Precision and Recall (Figure~\ref{fig:prec-recall-f1}). Lower thresholds increased the model's sensitivity to positive instances, improving Recall but also inflating the false positive rate. Conversely, higher thresholds suppressed spurious activations and improved Precision (up to 95\%), yet tended to under-detect true positives, reducing Recall. While a threshold of 0.6 offered a reasonable compromise between these metrics, we ultimately selected the standard 0.5 value to prioritize validation accuracy — the main criterion guiding our hyperparameter optimization.

The influence of BS was subtler, though relevant for convergence timing. For BS = 16, the average early stopping epoch was $33 \pm 14$ (median = 30), compared to $26 \pm 11$ (median = 24) for BS = 32, indicating that larger batch sizes may lead to more stable and efficient optimization trajectories. However, final validation accuracy remained comparable: $79.10\% \pm 10.93\%$ (BS = 16) versus $79.20\% \pm 11.08\%$ (BS = 32). This finding supports the use of smaller batch sizes when operating under hardware constraints, without sacrificing detection quality.

Our final selected configuration (BS = 32, $\eta = 10^{-4}$, threshold = 0.5) offers a robust operational balance and serves as the reference for subsequent fine-tuning and transfer learning experiments.

\subsection{AudioSet-EV Ablation Study}
We leveraged prior hyperparameter knowledge to conduct a progressively structured ablation study, assessing the isolated and cumulative impact of model- and training-level optimization strategies.

In the $1^{\text{st}}$ stage, only the final embedding and classification layers of the EPANNs were finetuned, while the rest of the network remained frozen. Training used a batch size of 32 and a learning rate $\eta=10^{-3}$, decayed to $10^{-6}$ over 10 epochs, with early stopping triggered after 33 stagnant epochs. Validation accuracy improved steadily from 80.1\% to 85.2\% by epoch 30. At test time, this partially finetuned model reached an accuracy of 84.2\%, with an $F_1$ score of 83.7\%, high Precision (91.4\%), and Recall at 77.2\%, indicating a configuration optimized for true positive detection. Total training time: 26.7 minutes.

The $2^{\text{nd}}$ experiment activated full backpropagation across all layers. Validation accuracy quickly peaked at 86\% by epoch 3, after which no further improvement was observed. The fully trained model significantly outperformed the previous one, reaching 95.9\% accuracy, 96.1\% $F_1$, Precision of 96.6\%, and Recall of 95.6\%, with both AuROC and AuPRC exceeding 95\%. Full training exploited the representational depth of the architecture and converged in 37.2 minutes (+10.5\%).

In the $3^{\text{rd}}$ stage, we retained full training but introduced a learning rate warm-up, linearly increasing $\eta$ from $10^{-6}$ to $10^{-3}$ over the first 5 epochs, followed by exponential decay for up to 100 epochs. Early stopping used a patience of 50. Validation accuracy rapidly reached 84.4\% (epoch 2), then 85.7\% (epoch 6). This led to our top-performing model on AudioSet-EV: 97.2\% accuracy, 97.1\% $F_1$, with Precision and Recall of 97.4\% and 96.8\%, respectively, and AuROC and AuPRC both exceeding 98\%. The warm-up phase smoothed early learning dynamics and enabled stable convergence. Training time: 57 minutes (GPU-accelerated).

The $4^{\text{th}}$ configuration extended the learning rate strategy with a cyclical cosine annealing schedule ($T_{\max}=100$ epochs), repeating for up to 1000 epochs with a patience of 150. This approach aimed to escape local minima via periodic learning rate boosts. Validation and test performance were comparable to the warm-up stage, with 96.7\% accuracy, $F_1$ of 97.2\%, Precision of 97.6\%, Recall of 96.8\%, and test time metrics above 98\% for both AuROC and AuPRC. These results confirm that both strategies support strong generalization, with the cyclical schedule offering greater flexibility.

Stages $5^{\text{th}}$ and $6^{\text{th}}$ introduced data augmentation while maintaining the cosine annealing schedule. In Stage 5, augmentations were applied with 70\% probability, yielding 91.5\% accuracy, 91.9\% $F_1$, Precision of 96.3\%, and Recall of 87.9\%. Although slightly lower than prior stages, results suggest improved robustness to data variability. In Stage 6, we replaced cyclical scheduling with a fixed $\eta=10^{-4}$ and reduced augmentation probability to 0.5. Surprisingly, this simpler configuration outperformed all previous setups, achieving test accuracy of 97.6\%, $F_1$ of 97.5\%, Precision of 97.8\%, and Recall of 97.3\%, highlighting how fixed learning rates — when paired with controlled augmentations — can promote both stable convergence and superior generalization, especially in heterogeneous datasets.

In summary, our ablation study suggests that training the full E2PANNs architecture with early-phase learning rate warm-up or fixed $\eta$ scheduling provides optimal performance. While both warm-up and cosine strategies yield robust models, results favor fixed $\eta$ when combined with data augmentations, especially under diverse training conditions. Notably, early stopping was triggered at variable epochs across experiments; hence, we re-evaluated all best checkpoints using a standardized testing procedure (AudioSet-EV test split, Seed 42, Table~\ref{tab:finetuning_test_metrics}).

Despite good performance across main EV datasets (Table~\ref{tab:finetuning_test_metrics}), we performed a final finetuning round on the best model from Stage 6, leveraging the entire EV benchmark ensemble (Section~\ref{sec:dataset_soa}, Table~\ref{tab:benchmark-partitions}) to validate its cross-dataset generalization.

\begin{table*}[h]
\centering
\caption{E2PANNs Test Metrics for each \textit{finetuning} stage}
\label{tab:finetuning_test_metrics}
\begin{threeparttable}
\scriptsize
\begin{tabular}{|c|c|c|c|c|c|c|c|c|c|c|}
\hline
\textbf{Finetuning Stage} & \textbf{Accuracy} & \textbf{Precision (PPV)} & \textbf{Recall (TPR)} & \textbf{F$_1$ Score} & \textbf{F$_2$ Score} & \textbf{Specificity (TNR)} & \textbf{AuROC} & \textbf{AuPRC} & \textbf{MCC} \\
\hline
1 & 87.3\% & 90.7\% & 83.8\% & 87.1\% & 88.0\% & 91.0\% & 87.4\% & 84.3\% & 74.9\% \\
2 & 88.7\% & \textbf{94.9\%} & 82.3\% & 88.1\% & 89.5\% & \textbf{95.3\%} & 88.8\% & 87.1\% & 78.1\% \\
3 & 89.0\% & 91.7\% & 86.2\% & 88.9\% & 89.5\% & 91.8\% & 89.0\% & 86.1\% & 78.1\% \\
4 & 89.0\% & 93.1\% & 84.7\% & 88.7\% & 89.6\% & 93.4\% & 89.1\% & 86.7\% & 78.3\% \\
5 & 87.5\% & 94.4\% & 80.4\% & 86.8\% & 88.4\% & 95.0\% & 87.7\% & 85.9\% & 76.0\% \\
\cellcolor{gray!30}6 & \cellcolor{gray!30}\textbf{90.5\%} & \cellcolor{gray!30}91.6\% & \cellcolor{gray!30}\textbf{89.7\%} & \cellcolor{gray!30}\textbf{90.6\%} & \cellcolor{gray!30}\textbf{90.9\%} & \cellcolor{gray!30}91.4\% & \cellcolor{gray!30}\textbf{90.5\%} & \cellcolor{gray!30}\textbf{87.4\%} & \cellcolor{gray!30}\textbf{81.1\%}\\
\hline
\noalign{\vspace{1.5em}}
\hline
\textbf{Stage 6 Benchmark} & \textbf{Accuracy} & \textbf{Precision (PPV)} & \textbf{Recall (TPR)} & \textbf{F$_1$ Score} & \textbf{F$_2$ Score} & \textbf{Specificity (TNR)} & \textbf{AuROC} & \textbf{AuPRC} & \textbf{MCC}\\
\hline
AudioSet-EV Augmented & \textbf{86.9\%} & \textbf{91.2\%} & \textbf{82.3\%} & \textbf{86.5\%} & \textbf{87.5\%} & \textbf{91.7\%} & \textbf{87.0\%} & \textbf{84.1\%} & \textbf{74.2\%} \\
\hline
\noalign{\vspace{0.5em}}
\hline
FSD50K & \textbf{94.6\%} & \textbf{16.3\%} & \textbf{69.1\%} & \textbf{26.4\%} & \textbf{23.2\%} & \textbf{94.9\%} & \textbf{82.0\%} & \textbf{11.7\%} & \textbf{31.9\%} \\
\hline
\noalign{\vspace{0.5em}}
\hline
LSSiren & \textbf{80.0\%} & \textbf{81.4\%} & \textbf{78.8\%} & \textbf{80.0\%} & \textbf{80.3\%} & \textbf{81.4\%} & \textbf{80.1\%} & \textbf{74.9\%} & \textbf{60.1\%} \\
\hline
\noalign{\vspace{0.5em}}
\hline
ESC50 Fold 1 & 87.5\% & 46.7\% & 87.5\% & 60.9\% & 57.1\% & 87.5\% & 87.5\% & 42.2\% & 58.0\% \\
ESC50 Fold 2 & 91.7\% & 58.3\% & 87.5\% & 70.0\% & 67.1\% & 87.5\% & 92.2\% & 52.4\% & 67.2\% \\
ESC50 Fold 3 & 95.8\% & 85.7\% & 75.0\% & 80.0\% & 81.2\% & 75.0\% & 98.4\% & 86.7\% & 77.9\% \\
ESC50 Fold 4 & 97.2\% & 100.0\% & 75.0\% & 85.7\% & 88.5\% & 75.0\% & 87.5\% & 77.8\% & 85.3\% \\
ESC50 Fold 5 & 90.3\% & 55.6\% & 62.5\% & 58.8\% & 58.1\% & 62.5\% & 93.8\% & 78.1\% & 53.5\% \\
\hline
\textit{Cross Validation} & \textbf{92.5\%} & \textbf{69.3\%} & \textbf{77.5\%} & \textbf{71.1\%} & \textbf{70.4\%} & \textbf{77.5\%} & \textbf{91.9\%} & \textbf{67.4\%} & \textbf{68.4\%} \\
\hline
\noalign{\vspace{0.5em}}
\hline
SirenNet Loader 0 & 100.0\% & 100.0\% & 100.0\% & 100.0\% & 100.0\% & 100.0\% & 100.0\% & 100.0\% & 100.0\% \\
SirenNet Loader 1 & 75.0\% & 75.0\% & 75.0\% & 75.0\% & 75.0\% & 75.0\% & 75.0\% & 68.8\% & 50.0\% \\
SirenNet Loader 2 & 87.5\% & 80.0\% & 100.0\% & 88.9\% & 86.8\% & 75.0\% & 87.5\% & 80.0\% & 77.5\% \\
SirenNet Loader 3 & 87.5\% & 80.0\% & 100.0\% & 88.9\% & 86.8\% & 75.0\% & 87.5\% & 80.0\% & 77.5\% \\
SirenNet Loader 4 & 92.4\% & 91.2\% & 93.9\% & 92.5\% & 92.2\% & 90.9\% & 92.4\% & 88.7\% & 84.9\% \\
SirenNet Loader 5 & 86.6\% & 79.5\% & 98.5\% & 88.0\% & 86.0\% & 74.6\% & 86.6\% & 79.1\% & 75.3\% \\
SirenNet Loader 6 & 86.6\% & 79.2\% & 99.3\% & 88.1\% & 86.0\% & 73.9\% & 86.6\% & 78.9\% & 75.6\% \\
SirenNet Loader 7 & 84.7\% & 77.7\% & 97.4\% & 86.4\% & 84.3\% & 72.0\% & 84.7\% & 77.0\% & 71.8\% \\
SirenNet Loader 8 & 87.0\% & 81.9\% & 98.7\% & 89.5\% & 87.7\% & 72.2\% & 85.5\% & 81.6\% & 75.1\% \\
SirenNet Loader 9 & 89.3\% & 87.5\% & 98.0\% & 92.4\% & 91.3\% & 72.2\% & 85.1\% & 87.1\% & 76.0\% \\
\hline
\textit{Average Results} & \textbf{87.7\%} & \textbf{83.2\%} & \textbf{96.1\%} & \textbf{89.0\%} & \textbf{87.6\%} & \textbf{78.1\%} & \textbf{87.1\%} & \textbf{82.1\%} & \textbf{76.4\%} \\
\hline
\noalign{\vspace{0.5em}}
\hline
US8K Fold 1 & 91.5\% & 53.8\% & 97.7\% & 69.4\% & 65.3\% & 90.9\% & 94.3\% & 52.8\% & 68.9\% \\
US8K Fold 2 & 84.5\% & 39.1\% & 92.3\% & 54.9\% & 50.4\% & 83.6\% & 87.9\% & 36.9\% & 53.7\% \\
US8K Fold 3 & 91.1\% & 59.9\% & 94.1\% & 73.2\% & 69.8\% & 90.7\% & 92.4\% & 57.1\% & 70.7\% \\
US8K Fold 4 & 93.5\% & 73.0\% & 97.6\% & 83.5\% & 80.9\% & 92.7\% & 95.2\% & 71.6\% & 80.9\% \\
US8K Fold 5 & 89.7\% & 41.3\% & 83.1\% & 55.1\% & 51.3\% & 90.3\% & 86.7\% & 35.6\% & 54.0\% \\
US8K Fold 6 & 92.3\% & 54.8\% & 85.1\% & 66.7\% & 63.6\% & 93.1\% & 89.1\% & 48.0\% & 64.5\% \\
US8K Fold 7 & 89.6\% & 45.5\% & 66.2\% & 53.9\% & 51.9\% & 91.9\% & 79.1\% & 33.3\% & 49.4\% \\
US8K Fold 8 & 92.3\% & 56.8\% & 93.8\% & 70.8\% & 67.1\% & 92.1\% & 92.9\% & 53.9\% & 69.4\% \\
US8K Fold 9 & 92.9\% & 58.7\% & 98.8\% & 73.6\% & 69.7\% & 92.2\% & 95.5\% & 58.1\% & 73.0\% \\
US8K Fold 10 & 84.5\% & 35.6\% & 69.9\% & 47.1\% & 44.0\% & 86.1\% & 78.0\% & 27.8\% & 42.2\% \\
\hline
\textit{Cross Validation} & \textbf{90.2\%} & \textbf{51.9\%} & \textbf{87.9\%} & \textbf{64.8\%} & \textbf{61.4\%} & \textbf{90.4\%} & \textbf{89.1\%} & \textbf{47.5\%} & \textbf{62.7\%} \\
\hline
\end{tabular}
\begin{tablenotes}
    \centering
      \footnotesize
      \item Upper Legend: \textbf{bold} -- best scores (per metric). Below the EV Benchmark for the average best model (Stage 6 Checkpoint).
    \end{tablenotes}
\end{threeparttable}
\end{table*}

\subsection{Unified-EV-Dataset Transfer Learning}
To improve E2PANNs generalization across diverse EV benchmark domains, we adopted a unified transfer learning strategy. The finetuning environment remained consistent with previous stages, except for GPU memory optimization, introduced to handle the larger volume of audio data. Memory allocation was improved by adopting CUDA expandable segments.

All audio files were re-sampled to a unified rate of 32kHz (aligned with prior finetuning stages), while original durations were preserved. Datasets were split according to their original benchmark specifications when available. Otherwise, we applied a standardized 80\%-10\%-10\% partitioning scheme, consistent with our AudioSet-EV protocol. Prepared partitions were then merged to form unified \texttt{loaders} for training, validation, and testing (Table~\ref{tab:benchmark-partitions}). To handle the inherent variability in input duration across datasets, a custom collation function dynamically padded input waveforms during batch loading to match the longest sequence in the batch. This approach ensures uniform batch structure while preserving temporal alignment. All scripts for dataset processing and training orchestration are publicly available [\href{https://github.com/StefanoGiacomelli/e2panns/tree/main/2_multi-dataset_finetuning}{GitHub}].

The transfer-learning experiment was performed using PyTorch Lightning, with GPU acceleration, checkpointing, and RT metric logging. Training parameters were: batch size = 16, fixed learning rate $\eta=10^{-4}$, $F_\beta$ with $\beta=2$, binary threshold = 0.5, and overall training enabled. The Adam optimizer was used with $\beta_1=0.9$, $\beta_2=0.999$, $\varepsilon=10^{-8}$, and weight decay $\lambda=10^{-6}$. Training was capped at 1000 epochs with an early stopping patience of 50.

\begin{figure}
    \centering
    \includegraphics[width=1\linewidth]{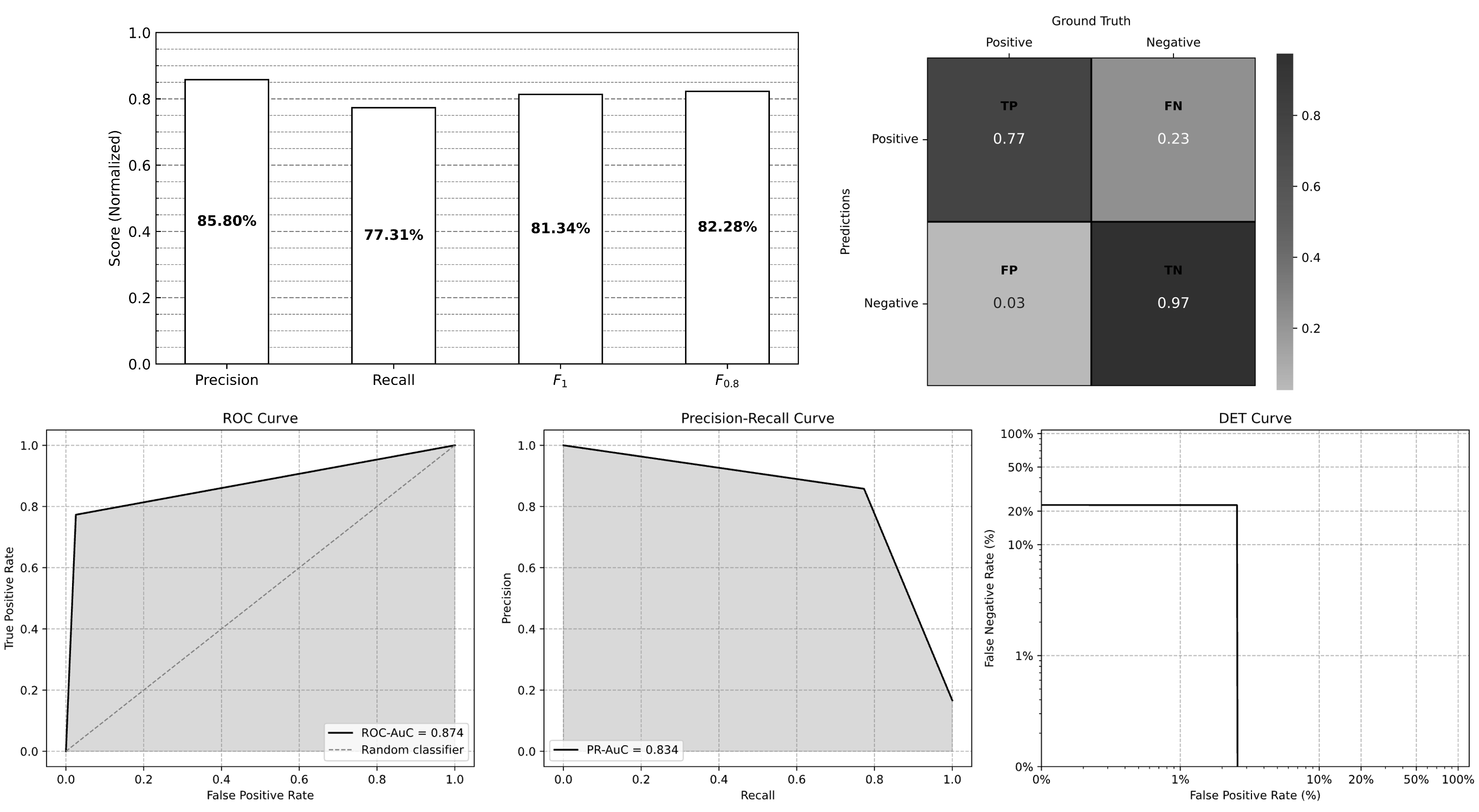}
    \caption{E2PANNs Unified-EV Dataset performance metrics.}
    \label{fig:unified_transfer_learning}
\end{figure}

Results from the transfer-learning stage (Figure~\ref{fig:unified_transfer_learning}, Table~\ref{tab:unified_metrics}) indicate that E2PANNs maintained high robustness and reliability in correctly identifying the absence of EV sirens (True Negative Accuracy: 97\%, Precision: 85.8\%). This property is critical in real-world scenarios — such as urban traffic management or autonomous vehicle alert systems — where avoiding false activations caused by acoustically similar events (e.g.: horns, alarms, or speech) is essential.

As anticipated, the model's True Positive accuracy declined when evaluated on the unified dataset relative to the individually trained benchmarks, yet it remained satisfactory at 77\% (AuROC: 87.4\%, Figure~\ref{fig:unified_transfer_learning}). When performance is re-analyzed per dataset (Table~\ref{tab:unified_metrics}), improvements are observed across all benchmarks, except for a slight drop on AudioSet-EV, likely due to its greater heterogeneity. Still, this reinforces the value of achieving strong performance on AudioSet-EV as a foundation for generalization to other datasets (Table~\ref{tab:finetuning_test_metrics}).

In conclusion, the unified strategy improves generalization and robustness over previous settings and compares favorably against existing SoA solutions (Section~\ref{soa}, Table~\ref{tab:ev_siren_comp}). To promote reproducibility and future benchmarking, we publicly release all pre-trained model checkpoints (\textcolor{red}{OneDrive Link}) that can be directly loaded in our inference-ready EPANNs re-implementation [\href{https://github.com/StefanoGiacomelli/epanns_inference}{GitHub}].

\begin{table*}[ht]
    \centering
    \caption{E2PANNs Benchmark post Unified-EV-Dataset Transfer Learning}
    \label{tab:unified_metrics}
    \scriptsize
    \begin{tabular}{|c|c|c|c|c|c|c|c|c|c|}
        \hline
        \textbf{Dataset} & \textbf{Accuracy} & \textbf{Precision (PPV)} & \textbf{Recall (TPR)} & \textbf{F$_1$ Score} & \textbf{F$_2$ Score} & \textbf{Specificity (TNR)} & \textbf{AuROC} & \textbf{AuPRC} & \textbf{MCC} \\
        \hline
        AudioSet-EV & \textbf{80.84\%} & \textbf{91.04\%} & \textbf{69.36\%} & \textbf{78.74\%} & \textbf{72.83\%} & \textbf{92.86\%} & \textbf{81.11\%} & \textbf{78.82\%} & \textbf{63.77\%} \\
        AudioSet-EV Augmented & \textbf{60.68\%} & \textbf{97.78\%} & \textbf{24.31\%} & \textbf{38.94\%} & \textbf{28.61\%} & \textbf{99.41\%} & \textbf{61.86\%} & \textbf{62.80\%} & \textbf{35.46\%} \\
        \hline
        \noalign{\vspace{0.5em}}
        \hline
        FSD50K & \textbf{98.81\%} & \textbf{61.11\%} & \textbf{40.00\%} & \textbf{48.35\%} & \textbf{42.97\%} & \textbf{99.64\%} & \textbf{69.82\%} & \textbf{25.28\%} & \textbf{48.88\%} \\
        \hline
        \noalign{\vspace{0.5em}}
        \hline
        LSSiren & \textbf{97.71\%} & \textbf{96.65\%} & \textbf{98.93\%} & \textbf{97.77\%} & \textbf{98.46\%} & \textbf{96.45\%} & \textbf{97.69\%} & \textbf{96.15\%} & \textbf{95.44\%} \\
        \hline
        \noalign{\vspace{0.5em}}
        \hline
        ESC50 Fold 1 & 100.00\% & 100.00\% & 100.00\% & 100.00\% & 100.00\% & 100.00\% & 100.00\% & 100.00\% & 100.00\% \\
        ESC50 Fold 2 & 100.00\% & 100.00\% & 100.00\% & 100.00\% & 100.00\% & 100.00\% & 100.00\% & 100.00\% & 100.00\% \\
        ESC50 Fold 3 & 95.83\% & 85.71\% & 75.00\% & 80.00\% & 76.92\% & 98.44\% & 86.72\% & 67.06\% & 77.90\% \\
        ESC50 Fold 4 & 98.61\% & 88.89\% & 100.00\% & 94.12\% & 97.56\% & 98.44\% & 99.22\% & 88.89\% & 93.54\% \\
        ESC50 Fold 5 & 100.00\% & 100.00\% & 100.00\% & 100.00\% & 100.00\% & 100.00\% & 100.00\% & 100.00\% & 100.00\% \\
        \hline
        \textit{Cross-Validation} & \textbf{98.89\%} & \textbf{94.92\%} & \textbf{95.00\%} & \textbf{94.82\%} & \textbf{94.90\%} & \textbf{99.38\%} & \textbf{97.19\%} & \textbf{91.19\%} & \textbf{94.29\%} \\
        \hline
        \noalign{\vspace{0.5em}}
        \hline
        SireNNet Loader 0 & 100.00\% & 100.00\% & 100.00\% & 100.00\% & 100.00\% & 100.00\% & 100.00\% & 100.00\% & 100.00\% \\
        SireNNet Loader 1 & 100.00\% & 100.00\% & 100.00\% & 100.00\% & 100.00\% & 100.00\% & 100.00\% & 100.00\% & 100.00\% \\
        SireNNet Loader 2 & 100.00\% & 100.00\% & 100.00\% & 100.00\% & 100.00\% & 100.00\% & 100.00\% & 100.00\% & 100.00\% \\
        SireNNet Loader 3 & 96.88\% & 100.00\% & 93.75\% & 96.77\% & 94.94\% & 100.00\% & 96.88\% & 96.88\% & 93.93\% \\
        SireNNet Loader 4 & 98.48\% & 100.00\% & 96.97\% & 98.46\% & 97.56\% & 100.00\% & 98.48\% & 98.48\% & 97.01\% \\
        SireNNet Loader 5 & 97.76\% & 98.48\% & 97.01\% & 97.74\% & 97.31\% & 98.51\% & 97.76\% & 97.04\% & 95.53\% \\
        SireNNet Loader 6 & 97.76\% & 99.23\% & 96.27\% & 97.73\% & 96.85\% & 99.25\% & 97.76\% & 97.39\% & 95.56\% \\
        SireNNet Loader 7 & 96.83\% & 99.22\% & 94.40\% & 96.75\% & 95.33\% & 99.25\% & 96.83\% & 96.46\% & 93.77\% \\
        SireNNet Loader 8 & 98.01\% & 99.43\% & 97.01\% & 98.21\% & 97.49\% & 99.29\% & 98.15\% & 98.13\% & 96.02\% \\
        SireNNet Loader 9 & 97.62\% & 99.63\% & 96.77\% & 98.18\% & 97.33\% & 99.29\% & 98.03\% & 98.56\% & 94.80\% \\
        \hline
        \textit{Average Results} & \textbf{98.33\%} & \textbf{99.60\%} & \textbf{97.22\%} & \textbf{98.38\%} & \textbf{97.68\%} & \textbf{99.56\%} & \textbf{98.39\%} & \textbf{98.29\%} & \textbf{96.66\%} \\
        \hline
        \noalign{\vspace{0.5em}}
        \hline
        US8K Fold 1 & 97.14\% & 85.06\% & 86.05\% & 85.55\% & 85.85\% & 98.35\% & 92.20\% & 74.56\% & 83.96\% \\
        US8K Fold 2 & 95.16\% & 77.91\% & 73.63\% & 75.71\% & 74.44\% & 97.62\% & 85.62\% & 60.06\% & 73.06\% \\
        US8K Fold 3 & 95.57\% & 89.80\% & 73.95\% & 81.11\% & 76.66\% & 98.76\% & 86.35\% & 69.76\% & 79.10\% \\
        US8K Fold 4 & 98.18\% & 93.53\% & 95.78\% & 94.64\% & 95.32\% & 98.67\% & 97.22\% & 90.29\% & 93.56\% \\
        US8K Fold 5 & 97.86\% & 88.06\% & 83.10\% & 85.51\% & 84.05\% & 99.08\% & 91.09\% & 74.46\% & 84.40\% \\
        US8K Fold 6 & 96.35\% & 89.29\% & 67.57\% & 76.92\% & 71.02\% & 99.20\% & 83.38\% & 63.24\% & 75.84\% \\
        US8K Fold 7 & 96.66\% & 91.53\% & 70.13\% & 79.41\% & 73.57\% & 99.34\% & 84.74\% & 66.93\% & 78.44\% \\
        US8K Fold 8 & 97.64\% & 86.75\% & 90.00\% & 88.34\% & 89.33\% & 98.48\% & 94.24\% & 79.06\% & 87.05\% \\
        US8K Fold 9 & 98.65\% & 93.83\% & 92.68\% & 93.25\% & 92.91\% & 99.32\% & 96.00\% & 87.70\% & 92.50\% \\
        US8K Fold 10 & 97.73\% & 88.10\% & 89.16\% & 88.62\% & 88.94\% & 98.67\% & 93.92\% & 79.62\% & 87.36\% \\
        \hline
        \textit{Cross-Validation} & \textbf{97.09\%} & \textbf{88.39\%} & \textbf{82.21\%} & \textbf{84.91\%} & \textbf{83.21\%} & \textbf{98.75\%} & \textbf{90.48\%} & \textbf{74.57\%} & \textbf{83.53\%} \\
        \hline
    \end{tabular}
\end{table*}

\subsection{Siren Recognition eXplainability}
To gain insights into the internal decision-making processes of our CNN-based EV siren classifier, we apply interpretability methods that visualize the model’s activation patterns and salient input features. Specifically, we employ two complementary eXplainable AI (XAI) techniques: Guided Backpropagation~\cite{SpringenbergDBR14}, a gradient-based approach highlighting the input spectrogram regions influencing predictions, and Score-CAM~\cite{9150840}, a gradient-free activation mapping method emphasizing the spatio-temporal features contributing to the model’s confidence. We present our analyses exploiting two representative siren samples randomly selected from the \textit{AudioSet-EV} test subset, both achieving inference probabilities above 95\%: a \textit{wail} siren, characterized by slow sinusoidal frequency modulation (period $\sim$4s/6s), and a \textit{yelp} siren, exhibiting rapid pitch modulations (period $\sim$0.5s).

\begin{figure}
    \centering
    \includegraphics[width=1\linewidth]{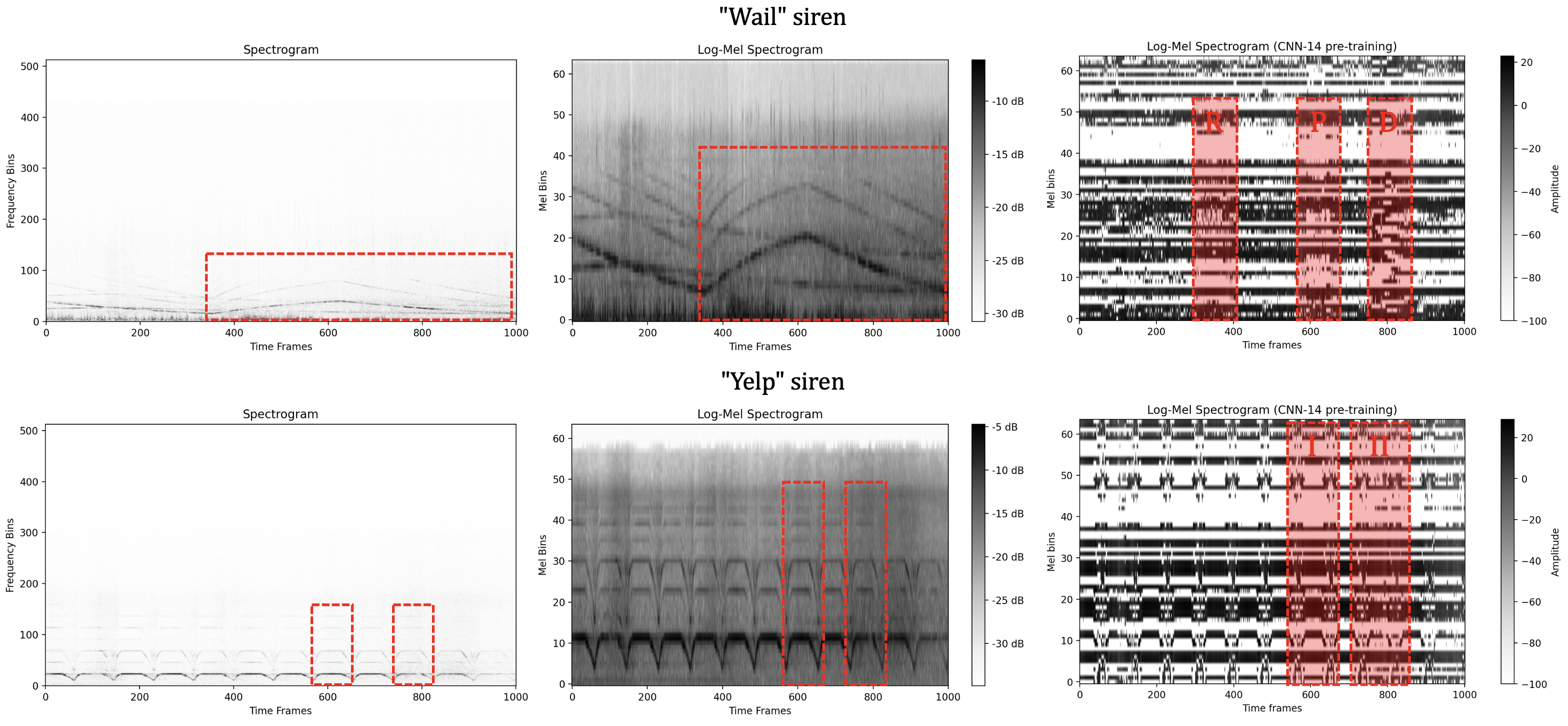}
    \caption{\textit{Left}: STFT modulus spectrograms extracted by the first embedded layer of E2PANNs (non trainable), \textit{Center}: Torchaudio Log-Mel spectrograms externally computed following E2PANNs hyperparameters, \textit{Right}: Log-Mel spectrograms computed by E2PANNs (CNN-14 pre-trained). \textit{Wail} siren: R - rising pitch phase, P - pitch peak, D - decay pitch phase; \textit{Yelp} siren: I - standard period, II - variant period.}
    \label{fig:spectrograms}
\end{figure}

The left column in Figure~\ref{fig:spectrograms} shows STFT spectrograms computed by the model’s initial fixed feature extraction layer. The central column displays standard log-mel spectrograms computed via Torchlibrosa~\cite{hwang_torchaudio_2023}, highlighting both foreground siren patterns and background urban noise. The \emph{wail} example demonstrates stronger reverberation indicative of greater source distance, whereas the \emph{yelp} example presents clearer acoustic signatures characteristic of closer proximity. The right column visualizes the internal log-mel spectrogram outputs from the E2PANNs model. Although this layer initially shared the same parameters as the standard log-mel extraction, differences emerge due to learned modifications from PANNs CNN-14 pre-training on AudioSet.

\begin{figure}
    \centering
    \includegraphics[width=1\linewidth]{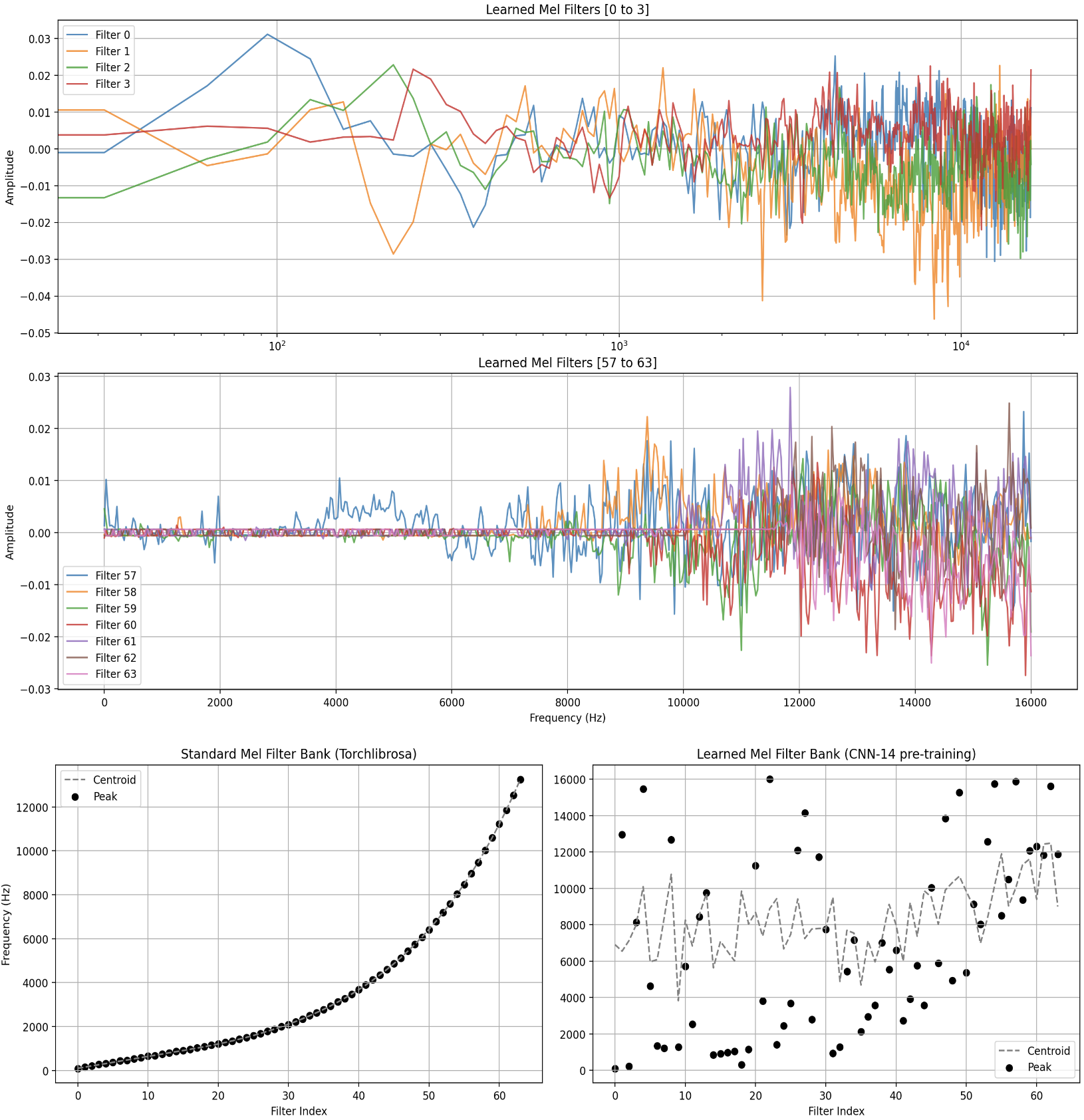}
    \caption{\textit{Top}: first four filters of the pre-trained E2PANNs log-mel filterbank, \textit{Center}: last seven filters of the pre-trained E2PANNs log-mel filterbank, \textit{Bottom-left}: initial Torchlibrosa Mel-filterbank centroids and peak frquencies, \textit{Bottom-right}: pre-trained Torchaudio Mel-filterbank centroids and peak frquencies from E2PANNs.}
    \label{fig:mel_filters}
\end{figure}

In particular, in Figure~\ref{fig:mel_filters}, we visualize the frequency-domain profiles of several learned mel filters from E2PANNs. As indicated by both peak frequency positions and frequency centroids, computed as:
\begin{equation}
\text{Centroid}_k = \frac{\sum_f f_c \cdot W_k(f)}{\sum_f W_k(f)},
\end{equation}
where $f$ denotes the frequency bin index, $f_c$ its corresponding center frequency (in Hz), and $W_k(f)$ the spectral weights of the $k$-th filter, the original log-spaced centroid distribution — typical of HTK-style triangular filter implementations~\cite{young2002htk} — is substantially altered by the PANNs AudioSet pre-training. The resulting filters lose their regular triangular shape, often including negative coefficients and broad side-lobes, indicative of reduced selectivity. While the centroid trend remains approximately pseudo-logarithmic, it appears compressed within a narrower frequency range (roughly 4kHz to 12kHz), and the peak response frequencies tend to drift far from centroids.

Although a coarse coverage of the full frequency spectrum (50Hz to 16kHz) is still preserved by the filterbank, Figure~\ref{fig:mel_filters} (upper plots) specifically highlights exemplifying filters affected by pre-training adaptations. These filters exhibit complex, non-monotonic frequency responses whose functional interpretability becomes evident only when correlated with the spectro-temporal feature extraction observed in Figure~\ref{fig:spectrograms}. There, one can identify the model’s capacity to extract multiple time-frequency variants of distinctive spectro-morphological siren patterns. Notably, the presence of replicated mid- and high-frequency features across filter channels suggests an emergent strategy aimed at enhancing time-frequency invariance, useful for robust detection under variable acoustic conditions.

These filter adaptations, inherited from the GP-AT multinomial pre-training phase, were not fine-tuned during our downstream learning (Table~\ref{tab:architecture_summary}). Nonetheless, they contribute meaningfully to the model’s discriminative capabilities in the EV siren recognition task, providing a structurally rich and semantically informative front-end representation that reinforces the inheritance of AT patterns learned during pre-training.

Guided Backpropagation~\cite{SpringenbergDBR14} is a saliency-based visualization method designed to highlight input features that most strongly influence the activation of a specific output neuron. It modifies standard backpropagation by suppressing negative gradients both in the ReLU backward pass and in the backpropagation process. Formally, Let $\phi(x) = \max(0, x)$ denote the ReLU activation function, applied element-wise to input $x$, and let $\delta_{\text{in}}$ denote the gradient flowing back from the upper layers. In standard backpropagation, the backward pass through a ReLU is defined as:
\begin{equation}
    \delta_{\text{out}} = \delta_{\text{in}} \cdot 1(x > 0),
\end{equation}
where $1(x > 0)$ is the indicator function, returning 1 only where the input $x$ is positive. This masks the gradient $\delta_{\text{in}}$ at all positions where the forward activation was zero, effectively stopping the gradient flow through ``inactive'' neurons. Guided Backpropagation adds an additional constraint further suppressing negative gradients in $\delta_{\text{in}}$, and allowing only positive signals to propagate backward. The modified rule becomes:
\begin{equation}
\delta_{\text{out}} = \delta_{\text{in}} \cdot 1(x > 0) \cdot 1(\delta_{\text{in}} > 0),
\end{equation}
where the second indicator function $1(\delta_{\text{in}} > 0)$ ensures that only positive gradients (those reinforcing the activation) are allowed to continue. We applied this method by backpropagating the class score gradients with respect to the input log-mel spectrogram, traversing the full CNNs stack. This resulted in saliency maps in the time-frequency domain that emphasize discriminative regions for the target class (322th in AudioSet). 

\begin{figure}
    \centering
    \includegraphics[width=1\linewidth]{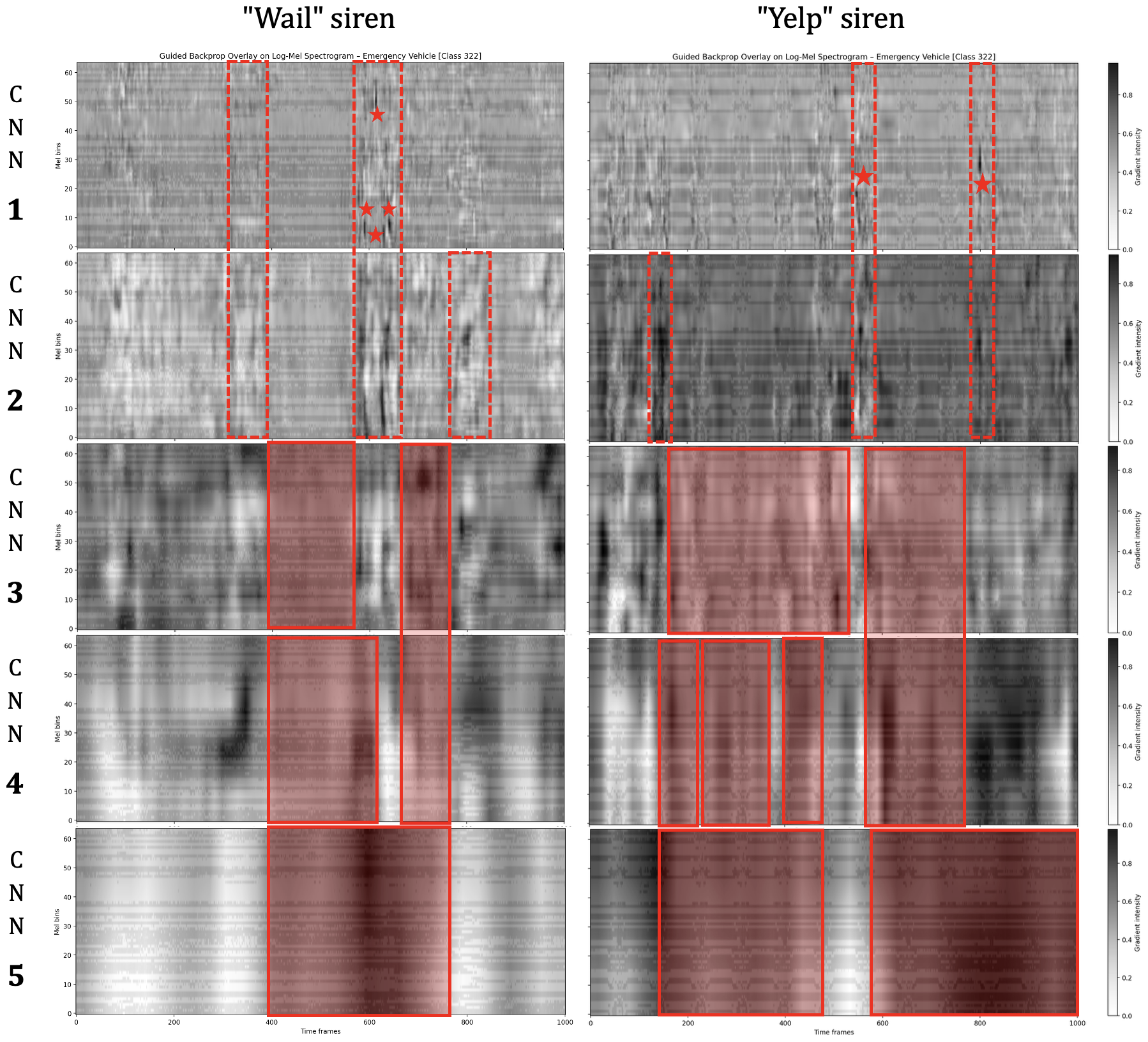}
    \caption{GuidedBackpropagation applied to the CNNs encoder of E2PANNs. Gradient intensity overlay on embedded Log-Mel spectrograms.}
    \label{fig:guidedbackprop}
\end{figure}

The layer-wise analysis (Figure~\ref{fig:guidedbackprop}) indicates a top-down abstraction approach: the first convolutional block (CNN 1) identifies time-energy anchors corresponding to distinct siren modulation phases—rising (R), peak (P), and decay (D) for \emph{wail}, and standard (I) and variant (II) cycles for \emph{yelp}. These patterns correspond to spectro-morphological energy blobs (marked by red-star tags): multiple and localized in \emph{wail}, singular and dispersed in \emph{yelp}. This suggests effective level of model specialization in detecting variable siren anchor patterns.

In CNN blocks 2 and 3, these features extend vertically, forming temporally robust anchors that demarcate boundaries of regions where pattern persistence (\emph{yelp}) or informational contiguity (\emph{wail}) confirms event presence. Subsequent convolutional blocks (CNN 3–5) progressively abstract and segregate features, becoming coarser in frequency and more decisive temporally, with gradient intensities varying horizontally. Gradient remapping onto input spectrograms required bilinear interpolation due to the convolutional encoder design, characterized by a progressive time-frequency bottleneck with redistribution across feature channels. Between CNN layers 3 and 4, the analysis transitions toward temporal redundancy, with CNN 5 summarizing information relative to overall siren pattern durations (highlighted regions correspond to contiguous medium-to-high gradient intensities). CNN 6 was excluded from the analysis, serving primarily as an informational reshaper before embedding and classification layers.

\begin{figure*}[ht]
    \centering
    \includegraphics[width=1\linewidth]{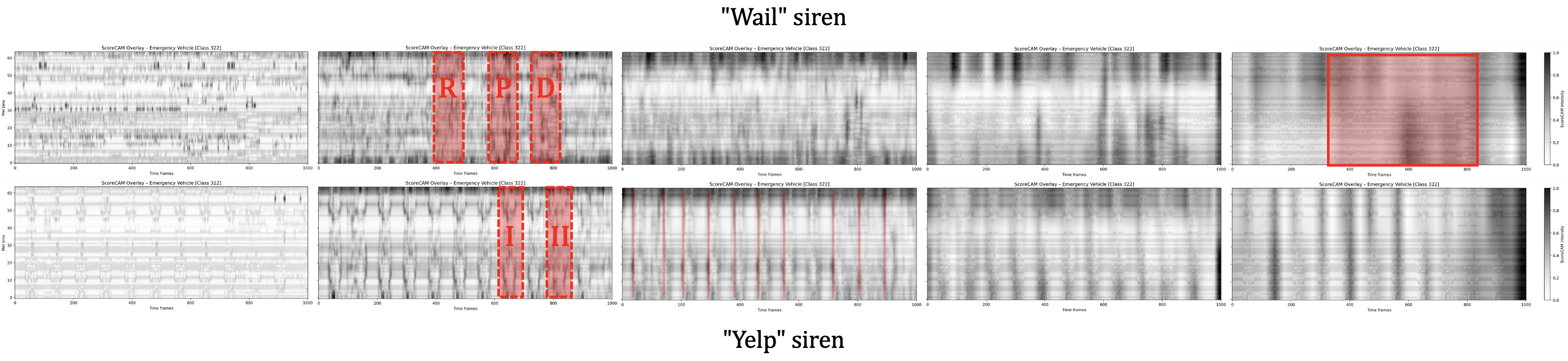}
    \caption{ScoreCAM applied to the E2PANNs CNNs encoder. From \textit{left} to \textit{right}: CNNs CAM intensity overlay on embedded Log-Mel spectrograms.}
    \label{fig:scorecam}
\end{figure*}

ScoreCAM~\cite{9150840} complements Guided Backpropagation by offering a gradient-free interpretability approach. Rather than relying on backpropagated gradients, ScoreCAM evaluates the contribution of intermediate activation maps to the model’s output score by masking the input with these activations and observing their influence on the predicted confidence. This provides robust and interpretable saliency maps, less sensitive to gradient noise and vanishing effects. Let $\hat{X}$ denote the input log-mel spectrogram, and let ${A_k}_{k=1}^{K}$ represent the set of activation maps from a target convolutional layer, with each $A_k \in \mathbb{R}^{H \times W}$ being the $k$-th feature map. These maps are first upsampled via bilinear interpolation to match the input dimensions, yielding $A_k^\uparrow \in \mathbb{R}^{bins \times frames}$. Each upsampled map is then normalized (min-max scaling) and element-wise multiplied with the input spectrogram to generate a masked input:
\begin{equation}
    M_k = M \cdot A_k^\uparrow
\end{equation}
This masked version $M_k$ is fed into the model, and a class-specific score $y_c(M_k)$ is computed, with $c=322$. This score reflects how relevant the $k$-th feature map is for predicting class $c$. To derive a normalized contribution weight for each activation map, a softmax normalization is applied across all $K$ maps:
\begin{equation}
    \alpha_k = \text{Softmax} \left(f_c(M \cdot A_k^\uparrow)\right) = \frac{\exp(f_c(M_k))}{\sum_j \exp(f_c(M_j))}.
\end{equation}
The final Class Activation Map (CAM) for class $c$ is then constructed as a linear combination of the up-sampled activation maps, weighted by their normalized class contributions:
\begin{equation}
    \text{CAM}_c = \sum{k=1}^K \alpha_k \cdot A_k^\uparrow.
\end{equation}
This map highlights regions in the time-frequency input domain that, when isolated, significantly contribute to increasing the class score $f_c$. As such, Score-CAM generates saliency maps that reflect causal relevance rather than mere correlation with the prediction. 

Layer-wise outputs (Figure~\ref{fig:scorecam}) confirm the model’s temporal aggregation of siren-specific patterns. Early convolutional stages (CNN 1–2) highlight detailed spectr0-temporal features essential for capturing modulation periodicities intrinsic to siren melodic profiles. By the second CNN layer, \emph{yelp} analysis already focuses on individual modulation cycles, while \emph{wail} remains focused on harmonic-rhythmic structures with emerging emphasis on R, P, and D intervals. Subsequent layers progressively cluster rhythmic-frequency anchors by similarity, ultimately forming higher-level periodic groupings. 

In summary, these analyses confirm that the E2PANNs model effectively leverages spectro-temporal pre-processing acquired during CNN-14 AT pre-training to robustly discriminate EV sirens, validating our architectural and fine-tuning choices and providing enhanced confidence in model interpretability.

\subsection{Embedded Real-Time Evaluation \& False Positive Analysis}
To evaluate the RT inference capabilities of the proposed E2PANNs model, we deployed it on a Raspberry Pi 5 (8 GB RAM, Broadcom BCM2712 2.4GHz quad-core 64-bit Arm Cortex-A76) using the custom multi-threaded inference pipeline described in Section~\ref{subsec:edge}. The test was performed on 1,025 positively labeled samples from the AudioSet-Strong dataset~\cite{Hershey2021TheBO} — a subset of AudioSet enhanced with temporally-aligned annotations — and used the best-performing checkpoint from our Unified-EV transfer learning phase (Table~\ref{tab:unified_metrics}).

Each 10-second sample was processed under simulated frame-wise RT inference conditions, with continuous logging of model outputs, frame durations, input sizes, and system-level performance metrics (CPU load, memory usage). The evaluation was repeated across three configurations for the inference window: one with fixed-length analysis frames, and two adaptive variants where the window width increased at rates of 0.2s and 0.4s in response to sustained confidence above threshold (0.5). This mechanism aimed to reduce false positives while preserving temporal sensitivity.

The preliminary analysis of model outputs revealed that 287 clips yielded no confident predictions (i.e.: no frames exceeded the 0.5 threshold). Manual audition of these clips showed that 182 (63.41\%) did not contain actual EV sounds, despite being positively labeled in AudioSet-Strong. This post-validation process exposed a significant presence of mis-annotated positives — referred to here as \textit{False} True Positives (FTP) — and led to the creation of a corrected metadata file that distinguishes valid EV samples from noisy annotations. This refinement strengthens the reliability of subsequent performance assessments and highlights the broader importance of dataset curation in safety-critical SED applications, where false alarms may lead to unacceptable outcomes.

To analyze detection quality under RT conditions, model outputs were evaluated using both Frame-Wise (FW) and Event-Based (EB) metrics. Frame-level probabilities were discretized into fixed-resolution binary sequences, aligned with the validated temporal annotations. This transformation was performed for both fixed and adaptive inference configurations, enabling compatibility with standard evaluation libraries: \texttt{torchmetrics}~\cite{detlefsen_torchmetrics_2022} for frame-wise metrics, and \texttt{sed\_eval}~\cite{app6060162} for event-based measures. The latter exploited the updated annotations and incorporated filtering of FTP-tagged clips, resulting in a stricter reference for scoring. Frame-wise metrics included Precision, Recall, F1-score, Accuracy, Specificity, Balanced Accuracy, and Error Rate. Event-based metrics included F-measure, Insertion and Deletion Rates, and Global Error Rate. Results are summarized in Table~\ref{tab:sed-evaluation-summary} for both FTP-filtered (True) and unfiltered (False) conditions.

\begin{table*}[ht]
    \centering
    \caption{Real-Time Evaluation summary for Unified-EV E2PANNs checkpoint}
    \label{tab:sed-evaluation-summary}
    \scriptsize
    \begin{threeparttable}
    \setlength{\tabcolsep}{3pt}
    \begin{tabular}{|c|c|c|c|c|c|c|c|c|c|c|c|c|c|}
    \hline
    \multirow{2}{*}{\textbf{FTP}} & \multirow{2}{*}{\makecell{\textbf{Framerate} \\ \textbf{Factor}}} &
    \multicolumn{6}{|c|}{\textbf{Frame-wise Metrics}} & 
    \multicolumn{3}{|c|}{\textbf{Event-based Accuracy}} & 
    \multicolumn{3}{|c|}{\textbf{Event-based Error Rate Metrics}} \\
    \cline{3-14}
    & & \textbf{Precision} & \textbf{Recall} & \textbf{F$_1$ score} & \textbf{Accuracy} & \textbf{Specificity} & \textbf{Bal. Accuracy} & 
    \textbf{F$_1$ score} & \textbf{Precision} & \textbf{Recall} & 
    \textbf{Error Rate} & \textbf{Deletion Rate} & \textbf{Insertion Rate} \\
    \hline
    True & None & 86\% & 60\% & 66\% & 61\% & 26\% & 61\% & 5\% & 3\% & 12\% & 4.37 & 0.88 & 3.49 \\
    True & +0.2s & 86\% & 61\% & 67\% & 62\% & 24\% & 62\% & 12\% & 9\% & 19\% & 2.82 & 0.81 & 2.01 \\
    \cellcolor{gray!30}True & \cellcolor{gray!30}+0.4s & \cellcolor{gray!30}86\% & \cellcolor{gray!30}63\% & \cellcolor{gray!30}68\% & \cellcolor{gray!30}64\% & \cellcolor{gray!30}24\% & \cellcolor{gray!30}64\% & \cellcolor{gray!30}14\% & \cellcolor{gray!30}10\% & \cellcolor{gray!30}21\% & \cellcolor{gray!30}2.60 & \cellcolor{gray!30}0.79 & \cellcolor{gray!30}1.81 \\
    \hline
    False & None & 77\% & 50\% & 56\% & 57\% & \textbf{32\%} & 57\% & 5\% & 3\% & 10\% & 3.99 & 0.90 & 3.09 \\
    False & +0.2s & 77\% & 52\% & 57\% & 58\% & 31\% & 58\% & 11\% & 8\% & 16\% & 2.65 & 0.84 & 1.81 \\
    \cellcolor{gray!30}False & \cellcolor{gray!30}+0.4s & \cellcolor{gray!30}77\% & \cellcolor{gray!30}53\% & \cellcolor{gray!30}58\% & \cellcolor{gray!30}59\% & \cellcolor{gray!30}30\% & \cellcolor{gray!30}59\% & \cellcolor{gray!30}12\% & \cellcolor{gray!30}10\% & \cellcolor{gray!30}17\% & \cellcolor{gray!30}2.48 & \cellcolor{gray!30}0.83 & \cellcolor{gray!30}1.65 \\
    \hline
    \end{tabular} 
    \begin{tablenotes}
      \centering
        \footnotesize
        \item \textbf{FTP} indicates whether evaluation excludes (\texttt{True}) or includes (\texttt{False}) suspected mis-annotations. 
        
        Accuracy Metrics: the higher - the better, Error Metrics: the lower - the better.
    \end{tablenotes}
    \end{threeparttable}
\end{table*}

To characterize the model’s robustness against spurious activations — an essential factor in low-latency safety-critical systems — we conducted a detailed analysis of false positive behaviors. A frame was considered a False Positive if its confidence score exceeded an improved threshold (0.6) without overlapping a ground truth event. Following large-scale SED evaluation conventions~\cite{Mesaros_2018, Cakir_2017}, we also analyzed false positive events: sequences of at least three consecutive FP frames. The metrics reported in Table~\ref{tab:FP_analysis} quantify both FW and EB false positive characteristics: average number of FP frames per sample (AFPS), percentage of false positives over total frames (AFPSP), average confidence of FP frames (AC), number of false positive events (FP), mean confidence of these events (AC), maximum run length (MRL), and average run length (ARL). Additionally, we report the average confidence (AC) across all inference frames for context. The results show that the \texttt{constant\_frame} configuration is more susceptible to persistent, high-confidence false activations, while both \texttt{variable\_frame} variants produce shorter and less frequent false positive bursts. This supports the hypothesis that adaptive windowing enhances robustness by leveraging temporally extended context to suppress spurious detections.

\begin{table}[ht]
    \centering
    \caption{False Positives Inference Analysis}
    \label{tab:FP_analysis}
    \scriptsize
    \begin{threeparttable}
    \setlength{\tabcolsep}{3pt}
    \begin{tabular}{|c|ccc|cccc|}
        \hline
        \multirow{2}{*}{\makecell{\textbf{Framerate} \\ \textbf{Factor}}} &
        \multicolumn{3}{c|}{\textbf{Frame-wise Metrics}} &
        \multicolumn{4}{c|}{\textbf{Event Based Metrics}} \\
        \cline{2-8}
        & \textbf{AFPS} & \textbf{AFPSP} & \textbf{AC} &
        \textbf{FP-TOT} & \textbf{AC} & \textbf{MRL} & \textbf{ARL} \\
        \hline
        None & 1.56 & 4.88\% & \textbf{58.73\%} & 116 & 84.76\% & 21 & 6.16 \\
        +0.2s & 0.64 & 3.94\% & 50.01\% & 58 & 84.36\% & 8 & 3.97 \\
        \cellcolor{gray!30}+0.4s & \cellcolor{gray!30}0.57 & \cellcolor{gray!30}3.97\% & \cellcolor{gray!30}50.45\% & \cellcolor{gray!30}53 & \cellcolor{gray!30}86.75\% & \cellcolor{gray!30}7 & \cellcolor{gray!30}3.85 \\
        \hline
    \end{tabular}
    \end{threeparttable}
\end{table}

System-level performance confirms the feasibility of embedded deployment. During active inference, the pipeline maintained a throughput of 1.35x the frame duration, with worst-case latency remaining under 400ms. In the idle condition, throughput matched the 1.00x input rate, with 99.53\% frame success and average CPU/RAM usage of 30.3\% and 15.5\%, respectively. No runtime interruptions were observed in either session. These findings confirm that E2PANNs can be effectively deployed as a full-stack SED solution on embedded platforms without requiring further low-level optimization.


\begin{table}[ht]
    \centering
    \caption{Server-grade Inference Profiling Results\\(Free Google Colab w. GPU acceleration)}
    \label{tab:profiling_results}
    \begin{threeparttable}
    \begin{tabular}{|c|c|}
        \hline
        \textbf{Metric} & \textbf{Value} \\
        \hline
        \multicolumn{2}{|c|}{CPU only}\\
        \hline
        Min. Input Size & 9919 samps (0.31s, 32KHz) \\
        CPU Benchmark Time & 0.218 $\pm$ 0.036s \\
        CPU Process Time & 0.822 $\pm$ 0.134s \\
        CPU Usage (Avg - Peak) & 50.63\% - 69.6\% \\
        CPU Energy (Avg) & 0.00026KWh\\
        \hline
        RAM Usage (Peak) & 0.098MB \\
        RAM Energy & 0.00012KWh \\
        \hline
        \multicolumn{2}{|c|}{GPU acceleration}\\
        \hline
        CUDA Kernel Time & 0.016 $\pm$ 0.048s \\
        GPU Memory (Peak) & 202.66MB \\
        GPU Usage (Avg - Peak) & 0\% - 0\% \\
        GPU Energy (Avg) & 0.00012KWh\\
        \hline
        \hline
        E2E (CPU+CUDA) Time & 0.0099 $\pm$ 0.0003s \\
        \hline
        CO2 emissions rate & 7.43 $\times 10^{-4}$g/s\\
        \hline
    \end{tabular}
    \begin{tablenotes}
    \centering
      \footnotesize
      \item Obtained through 100x runs (10s input): values in \textit{mean} $\pm$ \textit{std\_dev}, unless otherwise specified.
    \end{tablenotes}
    \end{threeparttable}
\end{table}

\begin{table*}[ht]
    \centering
    \caption{E2PANNs Architecture Summary}
    \label{tab:architecture_summary}
    \begin{tabular}{|l|c|c|c|c|c|c|c|}
        \hline
        \textbf{Layer} & \textbf{IN shape} & \textbf{OUT shape} & \textbf{Param \#} & \textbf{Param \%} & \textbf{Kernel Shape} & \textbf{MACs} & \textbf{Trainable} \\
        \hline
        Spectrogram (\texttt{Conv1D}) & [1, 320000] & [1, 1, 1001, 513] & 1,050,624 & 4.32\% & [1024] & 1,051,674,624 & False \\
        Log-Mel Filter-bank & [1, 1, 1001, 513] & [1, 1, 1001, 64] & 32,832 & 0.14\% & -- & -- & CNN-14 Pre-Trained \\
        BatchNorm 2D & [1, 64, 1001, 1] & [1, 64, 1001, 1] & 128 & -- & -- & 128 & True \\
        ConvBlock 1 & [1, 1, 1001, 64] & [1, 64, 500, 32] & 37,696 & 0.15\% & [3, 3] & 2,398,556,288 & True \\
        ConvBlock 2 & [1, 64, 500, 32] & [1, 128, 250, 16] & 221,696 & 0.91\% & [3, 3] & 3,539,200,000 & True \\
        ConvBlock 3 & [1, 128, 250, 16] & [1, 256, 125, 8] & 885,760 & 3.65\% & [3, 3] & 3,539,200,000 & True \\
        ConvBlock 4 & [1, 256, 125, 8] & [1, 256, 62, 4] & 1,180,672 & 4.86\% & [3, 3] & 1,179,648,000 & True \\
        ConvBlock 5 & [1, 256, 62, 4] & [1, 512, 31, 2] & 3,541,792 & 14.60\% & [3, 3] & 879,764,864 & True \\
        ConvBlock 6 & [1, 512, 31, 2] & [1, 1024, 31, 2] & 14,157,872 & 58.18\% & [3, 3] & 877,660,160 & True \\
        Linear FC (embedding) & [1, 1024] & [1, 2048] & 2,099,200 & 8.64\% & -- & 2,099,200 & True \\
        Linear FC (output) & [1, 2048] & [1, 527] & 1,079,823 & 4.45\% & -- & 1,079,823 & True \\
        \hline
        \multicolumn{8}{|c|}{
            \begin{tabular}{c}
                \textbf{TOT Params:} 24,289,295 \quad (\textit{Trained:} 23,205,839, \textit{Non-trained:} 1,083,456) \\
                \textbf{IN size:} 1.28 MB (10s) \quad \textbf{Fwd/Bwd:} 253.06 MB \quad \textbf{Model size:} 97.16 MB \\
                \textbf{MACs:} 13.52 $\times 10^{9}$ \quad \textbf{FLOPs:} 27.03 $\times 10^{9}$\\
                \textbf{Checkpoints:} \textcolor{red}{OneDrive Link} \quad \textbf{Repository:} https://github.com/StefanoGiacomelli/e2panns
            \end{tabular}
        } \\
        \hline
    \end{tabular}
\end{table*}

\section{Conclusions} \label{conclusions}
This paper introduced E2PANNs~\ref{tab:architecture_summary}, a lightweight CNN model derived from the EPANNs architecture and specialized for the RT detection of EV sirens in complex urban acoustic environments. Through extensive and rigorous experimentation, we demonstrated that the proposed system offers a compelling balance between classification accuracy, outliers rejection, computational efficiency, and hardware deployability on low-power edge devices.

Our work contributed along several axes. First, we released \textit{AudioSet-EV}~\cite{giacomelli-preprint}, a curated benchmark dataset derived from AudioSet with a focus on semantic consistency, binary label disambiguation, and transferability evaluation. The dataset enables reproducible training and testing for EV siren detection tasks and highlights nowadays limitations in the original AudioSet taxonomy, which we partially mitigated through metadata filtering, post-hoc validation and manual corrections. Second, we re-opted and refined an underrated pruned NNs model, through a selected and optimized set of training strategies and ablation studies, achieving substantial improvements over EV-SoA classification accuracies, particularly when evaluated across a Unified-EV benchmarking corpora (Table \ref{tab:unified_metrics}).

To further enhance E2PANNs generalization and avoid overfitting to dominant corpora, we are currently developing and testing a new promising learning strategy: \textit{sequential dataset-aware} training. This method processes each Unified-EV dataset in isolation, cycling through them in round-based epochs and dynamically adjusting their influence on the learning process based on observed validation improvements. Specifically, after training on a dataset for a fixed number of epochs, the model is validated not only on the same dataset but also across all other benchmarks. Based on the change in some reference metrics (\(\Delta F_1\)), adaptive weighting is applied: datasets with stagnating performance (\(\Delta F_1\) below a pre-defined threshold) are assigned a higher weight, encouraging the model to focus on under-performing domains. Conversely, datasets with already saturated performance (\(\Delta F_1\) above the threshold) are down-weighted to prevent over-optimization. These weights are applied directly in the training loss, scaling the Binary Cross-Entropy scores accordingly. This adaptive scheme also integrates early stopping and checkpointing per dataset, reducing unnecessary training and preserving the best-performing configuration for each domain. Preliminary tests suggest that this mechanism can effectively regularize training, promote cross-domain consistency, and guide the model toward improved robustness in real-world heterogeneous environments. Integration of this strategy into the main E2PANNs pipeline is currently under evaluation as part of future works.

In terms of system deployment, we implemented and validated a real-time inference pipeline capable of frame-wise classification and adaptive windowing~\cite{giordano_is2}. Results confirmed the feasibility of running E2PANNs on embedded platforms without further low-level optimization~\ref{tab:profiling_results}, achieving reliable inference latency and robustness under real-word computational constraints. The integration of \textit{Guided Backpropagation} and \textit{Score-CAM} further allowed us to interpret the learned representations, revealing that E2PANNs leverages high-level spectro-temporal features acquired during AudioSet pre-training, further specializing in multiple siren profiles detection. These mechanisms exposed filter-level adaptations and highlighted the model’s competence in segregating progressive acoustic anchors from auditory urban background noise.

Despite these strengths, several limitations remain. First, the training process does not currently include fine-tuning of the mel-filterbank layer, potentially limiting adaptability to domain-specific acoustic characteristics. We are going to repeat the finetuning ablation study including this layer updates and validating its standard-form and post-learning contribution to the overall performance. Second, although the false positive analysis demonstrated robustness to model's spurious activations, further efforts are required to improve confidence calibration — especially in safety-critical scenarios where false alarms can have disruptive effects. Third, the current inference pipeline is limited to single-channel monophonic inputs: extending support to multi-microphone arrays could enhance detection reliability under occlusion or noisy conditions, paving the way toward RT Direction-of-Arrival (DOA) estimation integration.

Moreover, while explainability techniques such as Guided Backpropagation and Score-CAM yielded valuable diagnostic insights, their integration into the training loop remains an open research direction. Enabling \textit{human-in-the-loop} workflows or confidence-aware alert systems that incorporate explainable acoustic evidence would greatly enhance transparency and trust in practical deployments. Lastly, from a deployment perspective, several optimizations remain to be explored. including: integer quantization of model parameters for efficient memory usage, evaluation in headless Linux environments to better reflect embedded use cases, and runtime acceleration via ONNX-based conversion pipelines to enable broader hardware portability and RT performance scaling. 

These directions aim to consolidate E2PANNs as a reliable building block in intelligent acoustic sensing pipelines for smart mobility and urban safety applications.

\ifCLASSOPTIONcaptionsoff
  \newpage
\fi



\bibliographystyle{IEEEtran}
\bibliography{./my_bib.bib}                   
%

%

\begin{IEEEbiographynophoto}{Stefano Giacomelli}
(Graduate Student Member, IEEE) was born in Rieti, Italy, in 1992. He received the diploma in electric guitar from the Saint Louis College of Music, Rome, Italy, in 2018. He received the B.A. degree (cum laude) in electroacoustic music composition in 2020 and the M.A. degree (cum laude and merit mention) in sound direction and technologies in 2023, both from the “A. Casella” Conservatory of Music (ConsAQ), L’Aquila, Italy.

Since 2023, he has been pursuing the Ph.D. degree in information and communication technologies with the Department of Information Engineering, Computer Science, and Mathematics (DISIM), University of L’Aquila (UnivAQ), L’Aquila, Italy. His research interests include digital signal processing, music information retrieval, and auditory scene analysis, with a focus on deep learning-based signal embedding and neural networks optimization, for which he has authored several papers. He actively collaborates on national and European interdisciplinary research projects bridging scientific and artistic domains, including musical extended realities and artistic/musicological research methodologies. He is also a composer, sound director, performer, and software developer for contemporary artistic productions.

Mr. Giacomelli is the recipient of several awards, including national prizes and merit mentions at the Italian National Arts Award (MUR).
\end{IEEEbiographynophoto}

\begin{IEEEbiographynophoto}{Marco Giordano}
(Graduate Student Member, IEEE) was born in Napoli, Italy, in 1965. He received the Laurea degree in electronic engineering from the University of Rome, Rome, Italy. He is currently pursuing the Ph.D. degree in information and communication technologies with the Department of Information Engineering, Computer Science, and Mathematics (DISIM), University of L’Aquila, L’Aquila, Italy.

He is a Full Professor of musical acoustics and perception at the “A. Casella” Conservatory of Music, L’Aquila, Italy. His research interests include digital signal processing, musical acoustics and perception, digital education, and deep neural networks applied to computational paralinguistics and sound recognition.
\end{IEEEbiographynophoto}

\begin{IEEEbiographynophoto}{Claudia Rinaldi}
(Member, IEEE) received the M.S. degree (cum laude) in telecommunication engineering from the University of L’Aquila, L’Aquila, Italy, in 2005, and the Ph.D. degree in electrical and information engineering from the same institution in 2009. She also received the Bachelor of Arts degree in electronic music in 2013 and the Master of Arts degree in trumpet in 2006, both from the “A. Casella” Conservatory of Music, L’Aquila, Italy.

From 2018 to 2022, she was a Researcher with the University of L’Aquila, focusing on digital signal processing, multimedia coding, and wireless communication systems. She is currently a Researcher (IV level) with the National Inter-University Consortium for Telecommunications (CNIT), L’Aquila, Italy, working on 5G-and-beyond networks, edge computing, and acoustic immersivity over networks. She is also an Adjunct Professor of digital and statistical signal processing and multimedia signal processing with programmable hardware design at the University of L’Aquila. She has contributed to several EU-funded and national research projects and is the author of numerous publications on wireless communications, multimedia compression, and signal processing techniques for the Internet of Things and cyber-physical systems.

Dr. Rinaldi is a member of the Center of Excellence DEWS Board of Directors at the University of L’Aquila. She has served as an Editor for journals such as Wireless Networks (Springer) and Sensors (MDPI).
\end{IEEEbiographynophoto}

\begin{IEEEbiographynophoto}{Fabio Graziosi}
(Senior Member, IEEE) received the Laurea degree in electronic engineering (telecommunications) in 1993 and the Ph.D. degree in electronic engineering in 1997, both from the University of L’Aquila, L’Aquila, Italy.

He is currently a Full Professor in telecommunications with the Department of Information Engineering, Computer Science, and Mathematics (DISIM), University of L’Aquila. His research interests include wireless communication systems, sensor networks, cognitive radio, cooperative communications, and cybersecurity. He has been involved in major national and European research programs and is the author of more than 200 publications in international journals and conference proceedings.

Prof. Graziosi has coordinated several large-scale projects, including post-earthquake smart city initiatives (Innovating City Planning Through Information and Communication Technologies – INCIPICT, L’Aquila) and 5G experimentation programs (Casa delle Tecnologie Emergenti – SICURA, L’Aquila). He is currently the President of the VITALITY Foundation, which leads the namesake innovation ecosystem funded under the Italian National Recovery and Resilience Plan (PNRR).
\end{IEEEbiographynophoto}




\end{document}